\documentclass[]{article}
\usepackage{lmodern}
\usepackage{amssymb,amsmath}
\usepackage{ifxetex,ifluatex}
\usepackage{fixltx2e} 
\ifnum 0\ifxetex 1\fi\ifluatex 1\fi=0 
  \usepackage[T1]{fontenc}
  \usepackage[utf8]{inputenc}
\else 
  \ifxetex
    \usepackage{mathspec}
  \else
    \usepackage{fontspec}
  \fi
  \defaultfontfeatures{Ligatures=TeX,Scale=MatchLowercase}
\fi
\IfFileExists{upquote.sty}{\usepackage{upquote}}{}
\IfFileExists{microtype.sty}{%
\usepackage{microtype}
\UseMicrotypeSet[protrusion]{basicmath} 
}{}
\usepackage[margin=1in]{geometry}
\usepackage{hyperref}
\hypersetup{unicode=true,
            pdftitle={An Introduction to Animal Movement Modeling with Hidden Markov Models using Stan for Bayesian Inference},
            pdfborder={0 0 0},
            breaklinks=true}
\usepackage{color}
\usepackage{fancyvrb}

\DefineVerbatimEnvironment{Highlighting}{Verbatim}{commandchars=\\\{\}}
\usepackage{framed}
\definecolor{shadecolor}{RGB}{248,248,248}
\newenvironment{Shaded}{\begin{snugshade}}{\end{snugshade}}
\newcommand{\KeywordTok}[1]{\textcolor[rgb]{0.13,0.29,0.53}{\textbf{#1}}}
\newcommand{\DataTypeTok}[1]{\textcolor[rgb]{0.13,0.29,0.53}{#1}}
\newcommand{\DecValTok}[1]{\textcolor[rgb]{0.00,0.00,0.81}{#1}}

\newcommand{\FloatTok}[1]{\textcolor[rgb]{0.00,0.00,0.81}{#1}}

\newcommand{\StringTok}[1]{\textcolor[rgb]{0.31,0.60,0.02}{#1}}

\newcommand{\CommentTok}[1]{\textcolor[rgb]{0.56,0.35,0.01}{\textit{#1}}}

\newcommand{\OtherTok}[1]{\textcolor[rgb]{0.56,0.35,0.01}{#1}}

\newcommand{\ControlFlowTok}[1]{\textcolor[rgb]{0.13,0.29,0.53}{\textbf{#1}}}
\newcommand{\OperatorTok}[1]{\textcolor[rgb]{0.81,0.36,0.00}{\textbf{#1}}}

\newcommand{\NormalTok}[1]{#1}
\usepackage{graphicx,grffile}
\makeatletter
\def\maxwidth{\ifdim\Gin@nat@width>\linewidth\linewidth\else\Gin@nat@width\fi}
\def\maxheight{\ifdim\Gin@nat@height>\textheight\textheight\else\Gin@nat@height\fi}
\makeatother
\setkeys{Gin}{width=\maxwidth,height=\maxheight,keepaspectratio}
\IfFileExists{parskip.sty}{%
\usepackage{parskip}
}{
\setlength{\parindent}{0pt}
\setlength{\parskip}{6pt plus 2pt minus 1pt}
}
\providecommand{\tightlist}{%
  \setlength{\itemsep}{0pt}\setlength{\parskip}{0pt}}
\ifx\paragraph\undefined\else
\let\oldparagraph\paragraph
\renewcommand{\paragraph}[1]{\oldparagraph{#1}\mbox{}}
\fi
\ifx\subparagraph\undefined\else
\let\oldsubparagraph\subparagraph
\renewcommand{\subparagraph}[1]{\oldsubparagraph{#1}\mbox{}}
\fi

\let\rmarkdownfootnote\footnote%
\def\footnote{\protect\rmarkdownfootnote}

\usepackage{titling}

  \title{An Introduction to Animal Movement Modeling with Hidden Markov Models
using Stan for Bayesian Inference}
  \pretitle{\vspace{\droptitle}\centering\huge}
  \posttitle{\par}
  \author{}
  \preauthor{}\postauthor{}
  \date{}
  \predate{}\postdate{}

\begin{document}
\maketitle

\textbf{Vianey Leos-Barajas}\(^1\) \& \textbf{Théo Michelot}\(^2\)

\begingroup\small
\(^1\)Iowa State University/Bielefeld University -
\href{mailto:vleosbarajas@wiwi.uni-bielefeld.de}{\nolinkurl{vleosbarajas@wiwi.uni-bielefeld.de}}

\(^2\)University of Sheffield -
\href{mailto:tmichelot1@sheffield.ac.uk}{\nolinkurl{tmichelot1@sheffield.ac.uk}}
\endgroup

\section{Introduction}\label{introduction}

Hidden Markov models (HMMs) are popular time series model in many fields
including ecology, economics and genetics. HMMs can be defined over
discrete or continuous time, though here we only cover the former. In
the field of movement ecology in particular, HMMs have become a popular
tool for the analysis of movement data because of their ability to
connect observed movement data to an underlying latent process,
generally interpreted as the animal's unobserved behavior. Further, we
model the tendency to persist in a given behavior over time.

Those already familiar with Michael Betancourt's case study
``Identifying Bayesian Mixture Models'' will see a natural extension
from the \textit{independent} mixture models that are discussed therein
to an HMM, which can also be referred to as a \textit{dependent} mixture
model. Notation presented here will generally follow the format of
Zucchini et al. (2016) and cover HMMs applied in an unsupervised case to
animal movement data, specifically positional data. We provide Stan code
to analyze movement data of the wild haggis as presented first in
Michelot et al. (2016). Implementing HMMs in Stan has also been covered
by Luis Damiano here: \url{https://github.com/luisdamiano/gsoc17-hhmm}
For a thorough overview of HMMs, see Zucchini et al. (2016).

\section{Hidden Markov Models}\label{hidden-markov-models}

An HMM is a doubly stochastic time series with an observed process
\((Y_t)\) that depends on an underlying state process \((S_t)\). The
observations \(\{Y_t\}_{t=1}^T\) are taken to be conditionally
independent given the states \(\{S_t\}_{t=1}^T\) and are generated by
\textit{so-called} state-dependent distributions, \(\{f_n\}_{n=1}^N\).
In this case we assume that \(S_t\) can take on a finite number
\(N \geq 1\) of states, such that we can also refer to this as an
\(N\)-state HMM. The evolution of states over time is governed by a
first-order Markov chain, i.e.~Pr(\(S_t| S_{t-1}, \ldots, S_1\)) =
Pr(\(S_t | S_{t-1}\)), with transition probability matrix
\(\boldsymbol{\Gamma}^{(t)} = \gamma_{i,j}^{(t)}\), where
\(\gamma_{i,j}^{(t)} =\) Pr(\(S_t=j | S_{t-1}=i\)) for
\(i,j=1, \ldots, N\). Assuming a time-homogeneous process, we have that
\(\boldsymbol{\Gamma}^{(t)} = \boldsymbol{\Gamma}\). A consequence of
this formulation is that the amount of time \(D_n\) spent in a given
state \(n\) (before switching to an other state) is a random variable
that follows a geometric distribution with parameter
\(1 - \gamma_{n,n}\), formally \(D_n \sim Geom(1-\gamma_{n,n})\) with
\(D_n \in \mathbb{N}\). Lastly, it is necessary to define the initial
state distribution \(\boldsymbol{\delta}^{(1)}\) for the state process
at time \(t=1\) with entries \(\delta^{(1)}_n\) = Pr(\(S_1 = n\)), for
\(n=1, \ldots, N\).

All together, an HMM is completely defined by specification of three
components:

\begin{itemize}
\tightlist
\item
  State-dependent distributions, \(\{f_n\}_{n=1}^N\)
\item
  Transition probability matrix,
  \(\boldsymbol{\Gamma}^{(t)} = \gamma_{i,j}^{(t)}\), for
  \(i,j=1, \ldots, N\)
\item
  Initial state distribution, \(\boldsymbol{\delta}^{(1)}\)

  \begin{itemize}
  \tightlist
  \item
    Stationary distribution,
    \(\boldsymbol{\delta} = \boldsymbol{\delta}\boldsymbol{\Gamma}\)
  \item
    Estimate the initial state distribution,
    e.g.~\(\boldsymbol{\delta}^{(1)} \sim \text{Dirichlet}(\boldsymbol{\nu})\)
  \end{itemize}
\end{itemize}

For a time-homogeneous process we can use the stationary distribution as
the initial state distribution, otherwise we can estimate the
distribution.

\subsection{Likelihood}\label{likelihood}

There are two functions referred to as the ``likelihood'' in the HMM
literature, the complete-data likelihood, i.e.~the joint distribution of
the observations and states, or the marginal likelihood, i.e.~the joint
distribution of the observations only. The complete-data likelihood is
written as follows,

\begin{equation}
f(\mathbf{y}, \mathbf{s}) = \mathcal{L}_c =  \delta^{(1)}_{s_1} \prod_{t=2}^T \gamma_{s_{t-1},s_t} \prod_{t=1}^T f_{s_t}(y_t) 
\end{equation}

The simplicity of the complete-data likelihood formulation may be one
reason why many conduct inference for parameters and states jointly,
typically through a Gibbs sampler, alternating between estimation of
states and parameters. In contrast, evaluation of the marginal
likelihood requires summation over all possible state sequences,

\begin{equation}
\mathcal{L}_m =  \sum_{s_1 = 1}^N \cdots \sum_{s_T = 1}^N \delta^{(1)}_{s_1} \prod_{t=2}^T \gamma_{s_{t-1},s_t} \prod_{t=1}^T f_{s_t}(y_t) 
\end{equation}

However, evaluation of the marginal likelihood is necessary for
implementation in Stan as the states are discrete random variables.
Zucchini et al. (2016) show that the marginal likelihood can be written
explicitly as a matrix product,

\begin{equation}
\mathcal{L}_m = \boldsymbol{\delta}^{(1)} \mathbf{P}(y_1) \boldsymbol{\Gamma} \mathbf{P}(y_2) \cdots \boldsymbol{\Gamma} \mathbf{P}(y_t) \mathbf{1}^{\top}
\end{equation}

for an \(N\times N\) matrix
\(\mathbf{P}(y_t) = \text{diag}\left(f_1(y_t), \ldots, f_N(y_t) \right)\)
and a vector of 1s of length\(N\), \(\mathbf{1} = (1, \ldots, 1)\). For
observations missing at random, we simply have
\(\mathbf{P}(y_t) = \mathbf{I}_{N\times N}\). The marginal likelihood
can be calculated efficiently with the forward algorithm, which
calculates the likelihood recursively. We define the forward variables
\(\boldsymbol{\alpha}_t\), beginning at time \(t=1\), as follows

\begin{equation}\label{like1} 
\boldsymbol{\alpha}_1 = \boldsymbol{\delta}^{(1)} \mathbf{P}(y_1), \qquad \boldsymbol{\alpha}_{t} = \boldsymbol{\alpha}_{t-1} \boldsymbol{\Gamma} \mathbf{P}(y_{t}), 
\end{equation}

Then, the marginal likelihood is obtained by summing over
\(\boldsymbol{\alpha}_T\),

\begin{equation}\label{like2}
\mathcal{L}_m = f(y_1,\ldots,y_T) =\sum_{i=1}^N \alpha_T(i) = \boldsymbol{\alpha}_T \boldsymbol{1}^\top.
\end{equation}

Notably, the computational effort involved in evaluating
\(\mathcal{L}_m\) is only linear in \(T\), the number of observations,
for a given number of states, \(N\). Direct evaluation of the likelihood
can result in numerical underflow. However, we can also use the forward
algorithm to evaluate the log marginal likelihood,
log(\(\mathcal{L}_m\)), and avoid underflow when calculating each
forward variable -- as demonstrated in the implementation in Stan model
given below.

For the HMM details we provide here, we assume the following:

\begin{itemize}
\tightlist
\item
  The state-dependent distributions are distinct,
  \(f_1 \neq \cdots \neq f_N\);
\item
  The t.p.m.~\(\boldsymbol{\Gamma}\) has full rank and is ergodic
\end{itemize}

These two points are sufficient for an HMM to be identifiable. The first
point is important when applying an HMM to animal movement data because
the states are assumed to reflect \emph{different} behaviors. The second
point indicates we would like for the animal to be able to transition
between behaviors across time.

\subsection{Priors}\label{priors}

An HMM has two main sets of parameters that require specification of
prior distributions, the parameters corresponding to the i)
state-dependent distributions and ii) transition probabilities, with a
possible third set if estimating the initial distribution as well.

However, because an HMM lies within the class of mixture models, the
lack of identifiability due to label-switching (i.e.~a reordering of
indices can lead to same joint distribution) should be taken into
account.

\subsubsection{State-dependent
distributions}\label{state-dependent-distributions}

As in the independent mixture models discussed in Betancourt (2017),
identification and inferences of the state-dependent distributions of an
HMM can be problematic. Issues related to label-switching can make it
difficult for the MCMC chains to efficiently explore the parameter
space. In practice, HMMs are also notorious for their multi-modality. As
such, some additional restrictions and information, such as ordering of
a subset of the parameters of interest and/or informative priors can aid
inference. For example, we can impose an ordering on the means,
\(\mu_1 < \mu_2 < \cdots < \mu_N\), of the state-dependent distributions
(if possible), which is easily done in Stan:

\begin{verbatim}
parameters {
  postive_ordered[N] mu;
}
\end{verbatim}

Other parametrizations can also be used to order the means. For example,
given \(\mu_1 \in \mathbb{R}\) and a vector of length \(N-1\),
\(\eta \in \mathbb{R}^+\), \(\mu_n = \mu_{n-1} + \eta_{n-1}\), for
\(n \in {2, \ldots, N}\).

\begin{verbatim}

parameters {
  real mu;
  vector<lower=0> etas[N-1];
}

transformed parameters{
  vector[N] ord_mus;
  
  ord_mus[1] = mu;
  for(n in 2:N)
   ord_mus[n] = ord_mus[n-1] + etas[n-1];
  
}
\end{verbatim}

As the state-dependent distributions reflect characteristics of the
observed data, priors for the parameters of interest should not place
the bulk of the probability on values that are unrealistic. Also, note
that because of potential label-switching, some type of ordering will
likely be needed so that the priors correspond to the appropriate
distributions (if not exchangeable). See Betancourt (2017) for similar
issues in mixture models.

\subsubsection{Transition Probability
Matrix}\label{transition-probability-matrix}

It is typically easier to form some type of intuition of the parameters
of the state-dependent distributions than entries of the t.p.m. However,
in animal movement data there is generally persistence in the estimated
states that we would like to capture (hence the reason for using HMMs).
For the model, this behavior corresponds to large diagonal entries,
\(\gamma_{n, n}\) for \(n \in \{1, \ldots, N\}\), typically
\textgreater{}0.8 in our own experience, though this could of course
vary depending on the temporal resolution of the data and question of
interest.

\section{Capturing Important Features of the Data and Model
Evaluation}\label{capturing-important-features-of-the-data-and-model-evaluation}

There are two features of the movement data that we aim to capture, a)
the marginal distribution of \(Y_t\) and b) the temporal dependence of
the observed data (e.g.~autocorrelation).

\subsection{\texorpdfstring{Marginal Distribution of \(Y_t\) and
Temporal
Dependence}{Marginal Distribution of Y\_t and Temporal Dependence}}\label{marginal-distribution-of-y_t-and-temporal-dependence}

The marginal distribution of \(y_t\) of an HMM is the distribution a
given observation at time \(t\) unconditional on the states,
\(f(y_t | \boldsymbol{\theta})\), with \(\boldsymbol{\theta}\)
reflecting the state-dependent parameters requiring estimation. For a
time-homogeneous process with stationary distribution,
\(\boldsymbol{\delta}\), the marginal distribution is derived as a
mixture of the state-dependent densities weighted by entries of
\(\boldsymbol{\delta}\),

\begin{equation*}
f(y_t | \boldsymbol{\theta}) = \delta_1 \cdot f_1(y_t) + \cdots + \delta_N \cdot f_N(y_t)
\end{equation*}

In the analysis of animal movement data, the stationary distribution can
give the ecologist an estimate of the proportion of time that the animal
exhibits the states (and related behaviors) overall. However, it is
important to not only report this result of the HMM because there are
inifitely many HMM formulations that lead to the same marginal
distribution for \(y_t\). For example,

\begin{table}[h!]
\centering
\begin{tabular}{l | c | c}
& HMM1 & HMM2 \\
\hline
State-dependent Dist. & $f_1 \sim N(0, 4)$; $f_1 \sim N(5, 5)$& $f_2 \sim N(0, 4)$; $f_2 \sim N(5, 5)$\\
TPM & $\boldsymbol{\Gamma} = \bigl( \begin{smallmatrix}0.7 &0.3 \\ 0.3 & 0.7\end{smallmatrix}\bigr)$ & $\boldsymbol{\Gamma} = \bigl( \begin{smallmatrix}0.95 &0.0.5 \\ 0.05 & 0.95\end{smallmatrix}\bigr)$ \\
Stationary Dist &$\boldsymbol{\delta} = \left( 0.5 \quad 0.5 \right)$&$\boldsymbol{\delta} = \left( 0.5 \quad 0.5 \right)$\\
Marginal Dist. & $0.5 \cdot f_1 + 0.5 \cdot f_2$&$0.5 \cdot f_1 + 0.5 \cdot f_2$
\end{tabular}
\end{table}

This result is a key difference between independent mixture models and
HMMs. An HMM is identifiable, even given the above result, because there
is dependence over time that we take into account via the transition
probability matrix, \(\boldsymbol{\Gamma}\). The marginal distribution
does not completely relay all of the information about the manner in
which the data were generated. In particular, taking into account the
temporal dependence, as an HMM does, allows for identification of
state-dependent distributions that may significantly overlap and other
flexible forms, see Alexandrovich et al. (2016) and Langrock et al.
(2015).

Aside from capturing the marginal distribution of \(y_t\), we also aim
to capture the temporal dependence present in the data. In particular,
the autocorrelation structure of data produced by the fitted HMM should
be comparable to the data itself. As a result, this can be a key
characteristics with which to do posterior predictive checking (Morales
et al., 2004).

\subsection{Assessing Model Adequacy Using Forecast (Pseudo-)Residuals
and Posterior Predictive
Checks}\label{assessing-model-adequacy-using-forecast-pseudo-residuals-and-posterior-predictive-checks}

\subsubsection{Forecast
(Pseudo-)Residuals}\label{forecast-pseudo-residuals}

One manner in which the fitted HMM can be assessed is through evaluation
of the (pseudo-)residuals. The pseudo-residuals are computed in two
steps. First, for continuous observations, the uniform pseudo-residuals
\(u_t\) are defined as

\begin{equation*}
  u_t = \Pr(Y_t \leq y_t | \mathbf{Y}^{(t-1)} = y^{(t-1)}),\quad t \in \{ 1, \dots, T \}.
\end{equation*}

Then, the normal pseudo-residuals are obtained as

\begin{equation*}
  r_t = \Phi^{-1} (u_t),\quad t \in \{ 1, \dots, T \},
\end{equation*}

where \(\Phi\) is the cumulative distribution function of the standard
normal distribution. If the fitted HMM is the true data-generating
process, the \(r_t\) have a standard normal distribution. In practice, a
qq-plot can be used to compare the distribution of pseudo-residuals to
the standard normal, and assess the fit. Further, the (pseudo-)residuals
of a fitted HMM should not be autocorrelated, indicating that the
dependence is adequately captured.

\subsubsection{Posterior Predictive
Checks}\label{posterior-predictive-checks}

Posterior predictive checks allow one to assess the adequacy of the
fitted model by generating \(M\) replicate data sets from the
distribution
\(f(\boldsymbol{y}^* \| \boldsymbol{y}) = f(\boldsymbol{y}^* | \boldsymbol{\theta})f(\boldsymbol{\theta} \| \boldsymbol{y})\).
In particular here, we use these checks to assess the fitted model's
ability to be interpreted as the data generating mechanism. The main
idea is that the model should be able to produce data that is similar to
the one observed in the key features defined a priori.

Given \(M\) posterior draws
\(\boldsymbol{\theta}_1^*, \ldots, \boldsymbol{\theta}_M^*\), we
generate M data sets from the distribution
\(f(\boldsymbol{y}^* \| \boldsymbol{\theta}^*)\). We then compare key
features of the replicate data sets to observed features. See Betancourt
(2018) for more details.

We demonstrate a few graphical posterior predictive checks in the HMM
examples.

\section{State Estimation}\label{state-estimation}

In animal movement modeling (the focus presented here), estimation of
the underlying state sequence is not the primary focus of the analysis
but rather a convenient byproduct of the HMM framework. It is most
important that the estimated state-dependent distributions can be
connected to biologically meaningful processes, though state estimation
can help one visualize the results of the fitted models.

There are two approaches to state estimation:

\begin{itemize}
\tightlist
\item
  Local State Decoding:
  \(Pr(S_t| y_1, \ldots, y_T, \boldsymbol{\theta})\) OR
\item
  Global State Decoding:
  \(Pr(S_1, \ldots, S_T | y_1, \ldots, y_T, \boldsymbol{\theta})\)
\end{itemize}

The first considers the distribution of the state at time \(t\),
\(S_t\), given the observations and estimated parameters
\(\boldsymbol{\theta}\). These distributions can be obtained through
implementation of the forward-backward algorithm.

The aim of the second approach is to obtain the most likely state
sequence given all of the observations. For this, we use the Viterbi
algorithm which returns the most likely state sequence given the
observations and estimated parameters. Both approaches are already
covered by Luis Damiano:
\url{https://github.com/luisdamiano/gsoc17-hhmm}. In general, both will
return similar (if not equal) results when it comes to state decoding
(assigning an observation to one of \(N\) states).

Going beyond assignment of observations to states and obtaining the
state probabilities at each point in time can also be highly
informative. In particular, two models may result in similar state
decodings yet correspond to different estimates of the parameters of
interest. While this is not a problem \emph{per se}, it can be difficult
to connect the estimated states to key biological processes when the
observations have large probabilities of being associated with more than
one state.

\section{Example: Fitting a 2-state
HMM}\label{example-fitting-a-2-state-hmm}

\begin{Shaded}
\begin{Highlighting}[]
\CommentTok{# Initialisation}
\KeywordTok{library}\NormalTok{(rstan)}
\KeywordTok{library}\NormalTok{(bayesplot)}
\KeywordTok{library}\NormalTok{(ggplot2)}
\KeywordTok{library}\NormalTok{(coda)}
\KeywordTok{library}\NormalTok{(circular)}
\KeywordTok{library}\NormalTok{(moveHMM)}

\KeywordTok{rstan_options}\NormalTok{(}\DataTypeTok{auto_write =} \OtherTok{TRUE}\NormalTok{)}
\KeywordTok{options}\NormalTok{(}\DataTypeTok{mc.cores =}\NormalTok{ parallel}\OperatorTok{::}\KeywordTok{detectCores}\NormalTok{())}
\NormalTok{pal <-}\StringTok{ }\KeywordTok{c}\NormalTok{(}\StringTok{"firebrick"}\NormalTok{,}\StringTok{"seagreen"}\NormalTok{,}\StringTok{"navy"}\NormalTok{) }\CommentTok{# colour palette}
\KeywordTok{set.seed}\NormalTok{(}\DecValTok{1}\NormalTok{)}
\end{Highlighting}
\end{Shaded}

Before getting into details about how HMMs are applied to animal
movement data, we present how to fit a basic HMM in Stan (Stan 2018). We
consider a \(2\)-state HMM with Gaussian state-dependent distributions
for the observation process \(X_t\). That is, at each time step
\(t=1,2,\dots\), we have

\begin{equation*}
Y_t \vert S_t = j \sim N(\mu_j,\sigma^2),
\end{equation*}

for \(j \in \{1,2\}\).

The following code simulates from the model. Note that here we take the
initial state distribution to be the stationary distribution.

\begin{Shaded}
\begin{Highlighting}[]
\CommentTok{# Number of states}
\NormalTok{N <-}\StringTok{ }\DecValTok{2}
\CommentTok{# transition probabilities}
\NormalTok{Gamma <-}\StringTok{ }\KeywordTok{matrix}\NormalTok{(}\KeywordTok{c}\NormalTok{(}\FloatTok{0.9}\NormalTok{,}\FloatTok{0.1}\NormalTok{,}\FloatTok{0.1}\NormalTok{,}\FloatTok{0.9}\NormalTok{),}\DecValTok{2}\NormalTok{,}\DecValTok{2}\NormalTok{)}
\CommentTok{# initial distribution set to the stationary distribution}
\NormalTok{delta <-}\StringTok{ }\KeywordTok{solve}\NormalTok{(}\KeywordTok{t}\NormalTok{(}\KeywordTok{diag}\NormalTok{(N)}\OperatorTok{-}\NormalTok{Gamma }\OperatorTok{+}\DecValTok{1}\NormalTok{), }\KeywordTok{rep}\NormalTok{(}\DecValTok{1}\NormalTok{, N))}
\CommentTok{# state-dependent Gaussian means}
\NormalTok{mu <-}\StringTok{ }\KeywordTok{c}\NormalTok{(}\DecValTok{1}\NormalTok{,}\DecValTok{5}\NormalTok{)}

\NormalTok{nobs <-}\StringTok{ }\DecValTok{1000}
\NormalTok{S <-}\StringTok{ }\KeywordTok{rep}\NormalTok{(}\OtherTok{NA}\NormalTok{,nobs)}
\NormalTok{y <-}\StringTok{ }\KeywordTok{rep}\NormalTok{(}\OtherTok{NA}\NormalTok{,nobs)}

\CommentTok{# initialise state and observation}
\NormalTok{S[}\DecValTok{1}\NormalTok{] <-}\StringTok{ }\KeywordTok{sample}\NormalTok{(}\DecValTok{1}\OperatorTok{:}\NormalTok{N, }\DataTypeTok{size=}\DecValTok{1}\NormalTok{, }\DataTypeTok{prob=}\NormalTok{delta)}
\NormalTok{y[}\DecValTok{1}\NormalTok{] <-}\StringTok{ }\KeywordTok{rnorm}\NormalTok{(}\DecValTok{1}\NormalTok{, mu[S[}\DecValTok{1}\NormalTok{]], }\DecValTok{2}\NormalTok{)}

\CommentTok{# simulate state and observation processes forward}
\ControlFlowTok{for}\NormalTok{(t }\ControlFlowTok{in} \DecValTok{2}\OperatorTok{:}\NormalTok{nobs) \{}
\NormalTok{  S[t] <-}\StringTok{ }\KeywordTok{sample}\NormalTok{(}\DecValTok{1}\OperatorTok{:}\NormalTok{N, }\DataTypeTok{size=}\DecValTok{1}\NormalTok{, }\DataTypeTok{prob=}\NormalTok{Gamma[S[t}\OperatorTok{-}\DecValTok{1}\NormalTok{],])}
\NormalTok{  y[t] <-}\StringTok{ }\KeywordTok{rnorm}\NormalTok{(}\DecValTok{1}\NormalTok{, mu[S[t]], }\DecValTok{2}\NormalTok{)}
\NormalTok{\}}

\KeywordTok{plot}\NormalTok{(y, }\DataTypeTok{col=}\NormalTok{pal[S], }\DataTypeTok{type=}\StringTok{"h"}\NormalTok{)}
\end{Highlighting}
\end{Shaded}

\begin{figure}

{\centering \includegraphics[width=0.7\linewidth]{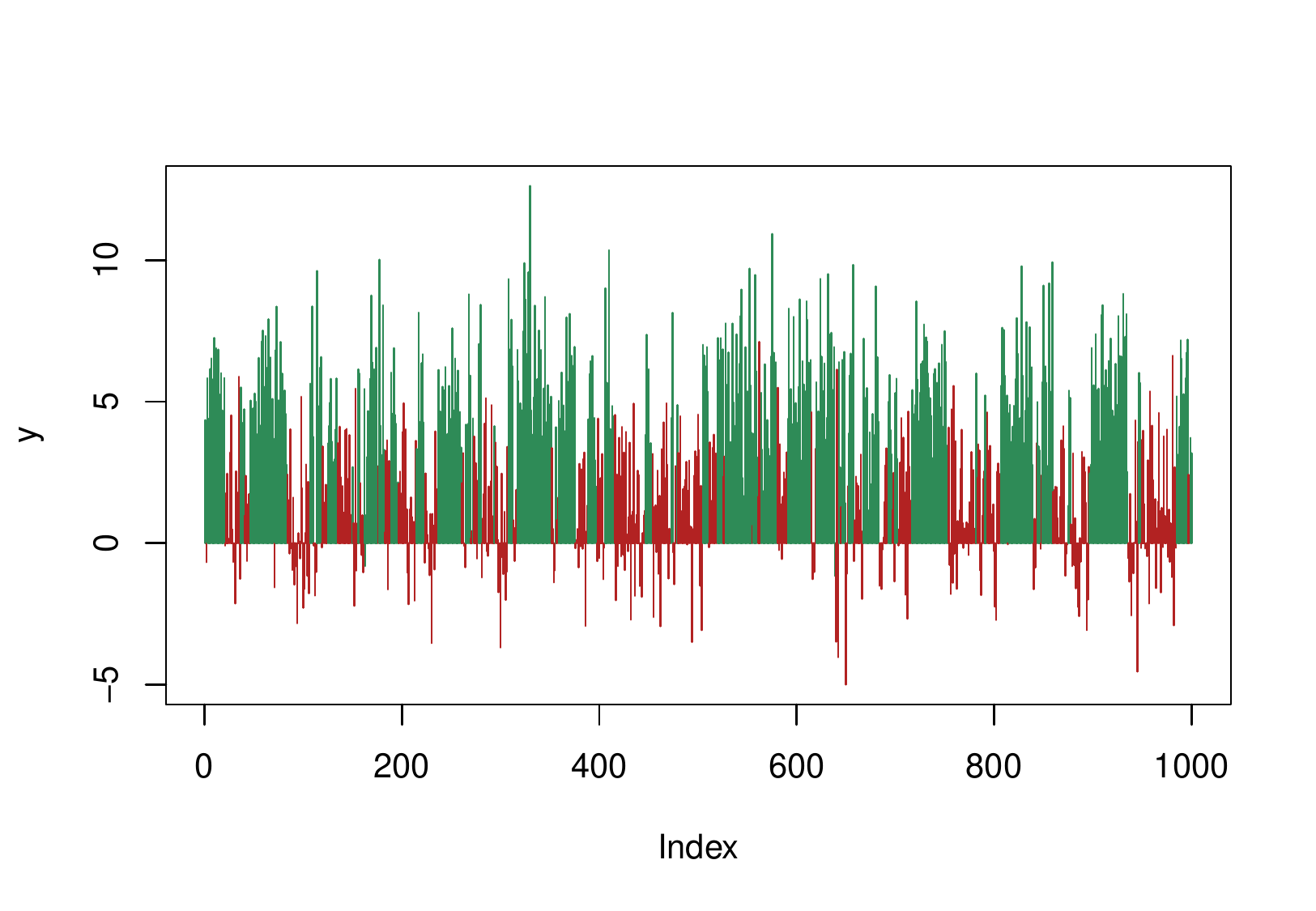} 

}

\caption{Simulated observations from a 2-state HMM with Gaussian state-dependent distributions.}\label{fig:unnamed-chunk-4}
\end{figure}

The likelihood of the model can be written with the forward algorithm,
given in Equation \ref{like1}, with

\begin{equation*}
P(x_t) =
\begin{pmatrix}
\phi(x_t \vert \mu_1, \sigma^2) & 0 \\
0 & \phi(x_t \vert \mu_2, \sigma^2) \\
\end{pmatrix},
\end{equation*}

where \(\phi\) is the Gaussian pdf.

The following code provides the complete implementation of the
\(N\)-state HMM with Gaussian state-dependent distributions in Stan,
based on the forward algorithm.

First, we define the known quantities in the data block:

\begin{verbatim}
data {
  int<lower=0> N; // number of states
  int<lower=1> T; // length of data set
  real y[T]; // observations
}
\end{verbatim}

There are two sets of parameters that requires estimation,
\(\boldsymbol{\Gamma}\) and \(\boldsymbol{\mu}\), which we define in the
parameter block:

\begin{verbatim}
parameters {
  simplex[N] theta[N]; // N x N tpm
  ordered[N] mu; // state-dependent parameters
}  
\end{verbatim}

We assume stationarity of the underlying Markov chain and initialize the
process with the stationary distribution, \(\boldsymbol{\delta}\). As
\(\boldsymbol{\delta}\) is a function of \(\boldsymbol{\Gamma}\), we
compute it in the transformed parameters block:

\begin{verbatim}
transformed parameters{
  
  matrix[N, N] ta; // 
  simplex[N] statdist; // stationary distribution
    
  for(j in 1:N){
    for(i in 1:N){
      ta[i,j]= theta[i,j];
    }
  }
  
  statdist =  to_vector((to_row_vector(rep_vector(1.0, N))/
      (diag_matrix(rep_vector(1.0, N)) - ta + rep_matrix(1, N, N)))) ;
}
\end{verbatim}

Given the information in the data block, and having defined all
parameters of interest, we now define the rest of the model in the model
block:

\begin{verbatim}
model {

  vector[N] log_theta_tr[N];
  vector[N] lp;
  vector[N] lp_p1;
  
  // prior for mu
  mu ~ student_t(3, 0, 1);

  // transpose the tpm and take natural log of entries
  for (n_from in 1:N)
  for (n in 1:N)
    log_theta_tr[n, n_from] = log(theta[n_from, n]);

  
  // forward algorithm implementation
  
  for(n in 1:N) // first observation
    lp[n] = log(statdist[n]) + normal_lpdf(y[1] | mu[n], 2);

  for (t in 2:T) { // looping over observations
    for (n in 1:N) // looping over states
      lp_p1[n] = log_sum_exp(log_theta_tr[n] + lp) + 
        normal_lpdf(y[t] | mu[n], 2); 
      
    lp = lp_p1;
  }

  target += log_sum_exp(lp);
}
\end{verbatim}

We first run 2000 iterations for each of the 4 chains, with the first
1000 draws drawn during the warm-up phase, and verify that the posterior
draws capture the true parameters.

\begin{Shaded}
\begin{Highlighting}[]
\NormalTok{stan.data <-}\StringTok{ }\KeywordTok{list}\NormalTok{(}\DataTypeTok{y=}\NormalTok{y, }\DataTypeTok{T=}\NormalTok{nobs, }\DataTypeTok{N=}\DecValTok{2}\NormalTok{)}
\NormalTok{fit <-}\StringTok{ }\KeywordTok{stan}\NormalTok{(}\DataTypeTok{file=}\StringTok{"HMM1.stan"}\NormalTok{, }\DataTypeTok{data=}\NormalTok{stan.data, }\DataTypeTok{refresh=}\DecValTok{2000}\NormalTok{)}
\end{Highlighting}
\end{Shaded}

\begin{Shaded}
\begin{Highlighting}[]
\NormalTok{mus <-}\StringTok{ }\KeywordTok{extract}\NormalTok{(fit, }\DataTypeTok{pars=}\KeywordTok{c}\NormalTok{(}\StringTok{"mu"}\NormalTok{))}

\KeywordTok{hist}\NormalTok{(mus[[}\DecValTok{1}\NormalTok{]][,}\DecValTok{1}\NormalTok{],}\DataTypeTok{main=}\StringTok{""}\NormalTok{,}\DataTypeTok{xlab=}\KeywordTok{expression}\NormalTok{(mu[}\DecValTok{1}\NormalTok{]))}
\KeywordTok{abline}\NormalTok{(}\DataTypeTok{v=}\DecValTok{1}\NormalTok{, }\DataTypeTok{col=}\NormalTok{pal[}\DecValTok{1}\NormalTok{], }\DataTypeTok{lwd=}\DecValTok{2}\NormalTok{)}
\KeywordTok{hist}\NormalTok{(mus[[}\DecValTok{1}\NormalTok{]][,}\DecValTok{2}\NormalTok{],}\DataTypeTok{main=}\StringTok{""}\NormalTok{,}\DataTypeTok{xlab=}\KeywordTok{expression}\NormalTok{(mu[}\DecValTok{2}\NormalTok{]))}
\KeywordTok{abline}\NormalTok{(}\DataTypeTok{v=}\DecValTok{5}\NormalTok{, }\DataTypeTok{col=}\NormalTok{pal[}\DecValTok{2}\NormalTok{], }\DataTypeTok{lwd=}\DecValTok{2}\NormalTok{)}
\end{Highlighting}
\end{Shaded}

\begin{figure}
\includegraphics[width=0.49\linewidth]{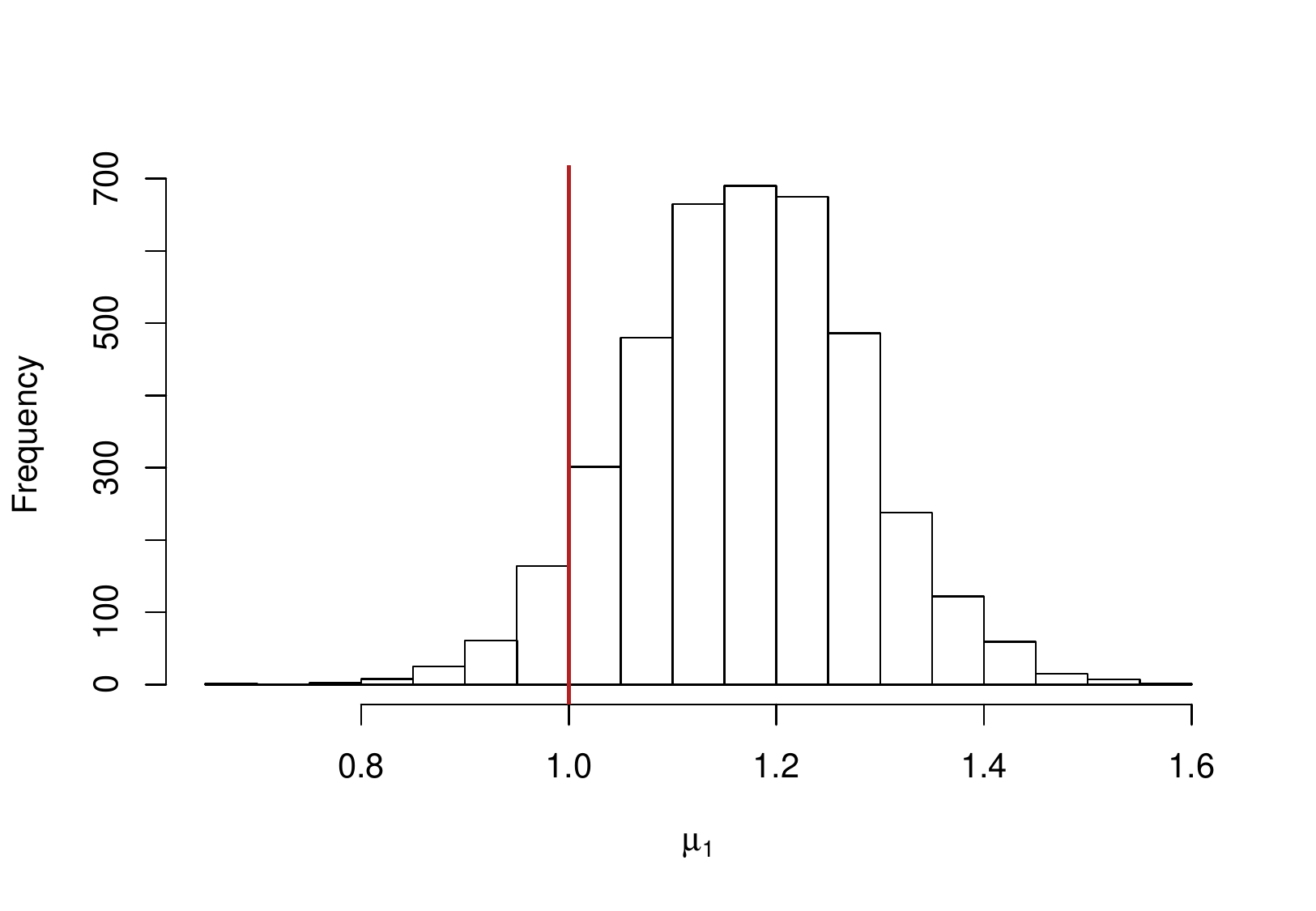} \includegraphics[width=0.49\linewidth]{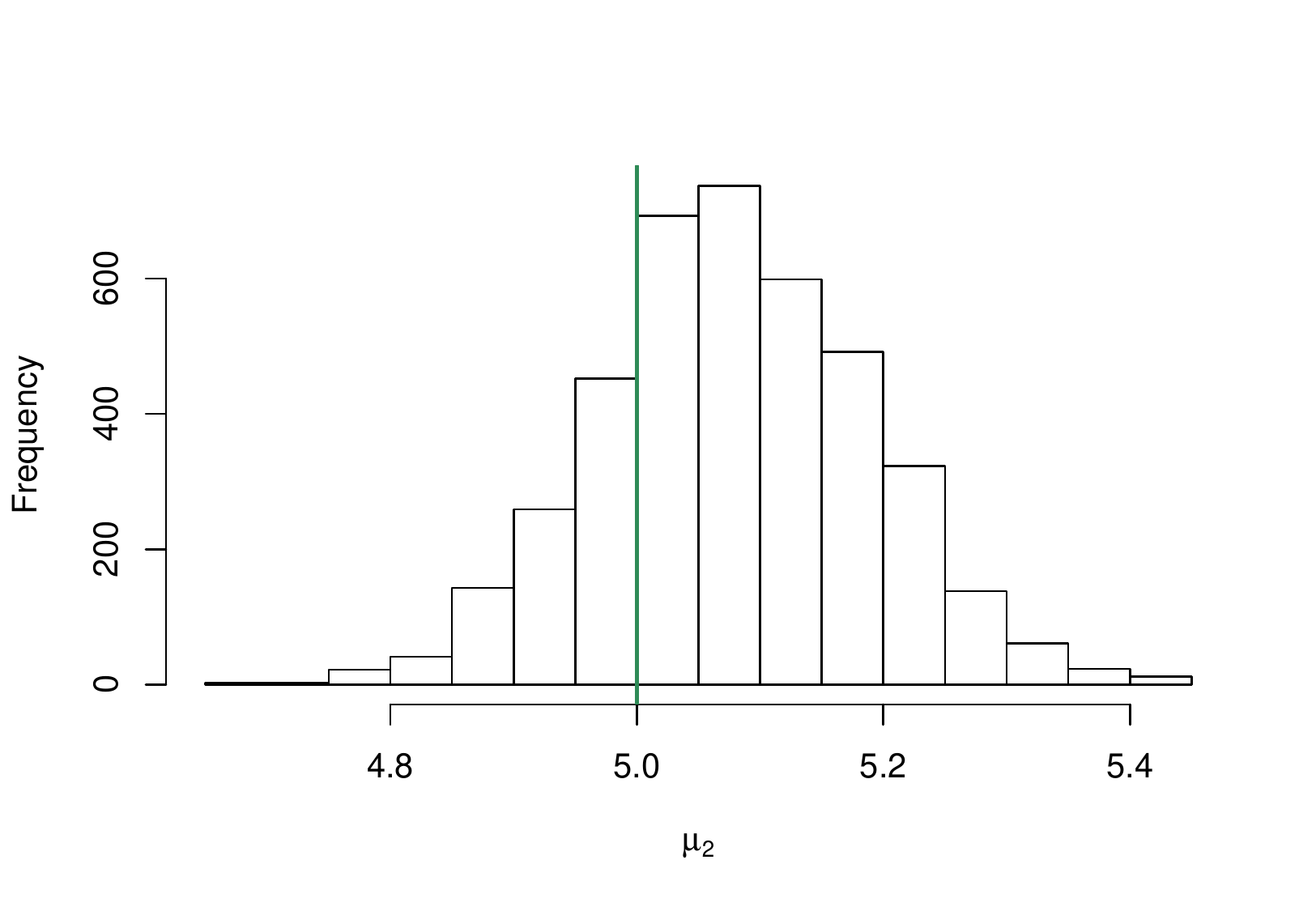} \caption{Histograms of the posterior draws for the state-dependent variance. The vertical lines show the true values used in the simulation.}\label{fig:unnamed-chunk-10}
\end{figure}

From the fitted model, we extract the parameters of interest and
generate 4000 data sets in order to perform a few graphical posterior
predictive checks.

\begin{Shaded}
\begin{Highlighting}[]
\NormalTok{## extract posterior draws}
\NormalTok{psam <-}\StringTok{ }\KeywordTok{extract}\NormalTok{(fit, }\DataTypeTok{pars =} \KeywordTok{c}\NormalTok{(}\StringTok{"theta"}\NormalTok{, }\StringTok{"mu"}\NormalTok{))}

\NormalTok{## generate new data sets}

\NormalTok{n.sims <-}\StringTok{ }\KeywordTok{dim}\NormalTok{(psam[[}\DecValTok{1}\NormalTok{]])[}\DecValTok{1}\NormalTok{]}
\NormalTok{n <-}\StringTok{ }\KeywordTok{length}\NormalTok{(y)}

\CommentTok{# state sequences}
\NormalTok{ppstates <-}\StringTok{ }\KeywordTok{matrix}\NormalTok{(}\OtherTok{NA}\NormalTok{, }\DataTypeTok{nrow =}\NormalTok{ n.sims, }\DataTypeTok{ncol =}\NormalTok{ n)}
\CommentTok{# observations}
\NormalTok{ppobs <-}\StringTok{ }\KeywordTok{matrix}\NormalTok{(}\OtherTok{NA}\NormalTok{, }\DataTypeTok{nrow =}\NormalTok{ n.sims, }\DataTypeTok{ncol =}\NormalTok{ n)}


\ControlFlowTok{for}\NormalTok{ (j }\ControlFlowTok{in} \DecValTok{1}\OperatorTok{:}\NormalTok{n.sims) \{}
\NormalTok{    theta <-}\StringTok{ }\NormalTok{psam[[}\DecValTok{1}\NormalTok{]][j, , ]}
\NormalTok{    statdist <-}\StringTok{ }\KeywordTok{solve}\NormalTok{(}\KeywordTok{t}\NormalTok{(}\KeywordTok{diag}\NormalTok{(N) }\OperatorTok{-}\StringTok{ }\NormalTok{theta }\OperatorTok{+}\StringTok{ }\DecValTok{1}\NormalTok{), }\KeywordTok{rep}\NormalTok{(}\DecValTok{1}\NormalTok{, N))}
    
\NormalTok{    ppstates[j, }\DecValTok{1}\NormalTok{] <-}\StringTok{ }\KeywordTok{sample}\NormalTok{(}\DecValTok{1}\OperatorTok{:}\NormalTok{N, }\DataTypeTok{size =} \DecValTok{1}\NormalTok{, }\DataTypeTok{prob =}\NormalTok{ statdist)}
\NormalTok{    ppobs[j, }\DecValTok{1}\NormalTok{] <-}\StringTok{ }\KeywordTok{rnorm}\NormalTok{(}\DecValTok{1}\NormalTok{, }\DataTypeTok{mean =}\NormalTok{ psam[[}\DecValTok{2}\NormalTok{]][j, ppstates[j, }\DecValTok{1}\NormalTok{]], }\DataTypeTok{sd =} \DecValTok{2}\NormalTok{)}
    
    \ControlFlowTok{for}\NormalTok{ (i }\ControlFlowTok{in} \DecValTok{2}\OperatorTok{:}\KeywordTok{length}\NormalTok{(y)) \{}
\NormalTok{        ppstates[j, i] <-}\StringTok{ }\KeywordTok{sample}\NormalTok{(}\DecValTok{1}\OperatorTok{:}\NormalTok{N, }\DataTypeTok{size =} \DecValTok{1}\NormalTok{, }\DataTypeTok{prob =}\NormalTok{ theta[ppstates[j, i }\OperatorTok{-}\StringTok{ }
\StringTok{            }\DecValTok{1}\NormalTok{], ])}
\NormalTok{        ppobs[j, i] <-}\StringTok{ }\KeywordTok{rnorm}\NormalTok{(}\DecValTok{1}\NormalTok{, }\DataTypeTok{mean =}\NormalTok{ psam[[}\DecValTok{2}\NormalTok{]][j, ppstates[j, i]], }\DataTypeTok{sd =} \DecValTok{2}\NormalTok{)}
\NormalTok{    \}}
\NormalTok{\}}
\end{Highlighting}
\end{Shaded}

First, we check that the densities of the replicated data sets are
similar to the observed data set. For this we use the R package
\texttt{bayesplot}.

\begin{Shaded}
\begin{Highlighting}[]
\KeywordTok{ppc_dens_overlay}\NormalTok{(y, ppobs[}\DecValTok{1}\OperatorTok{:}\DecValTok{100}\NormalTok{,])}
\end{Highlighting}
\end{Shaded}

\begin{center}\includegraphics[width=0.7\linewidth]{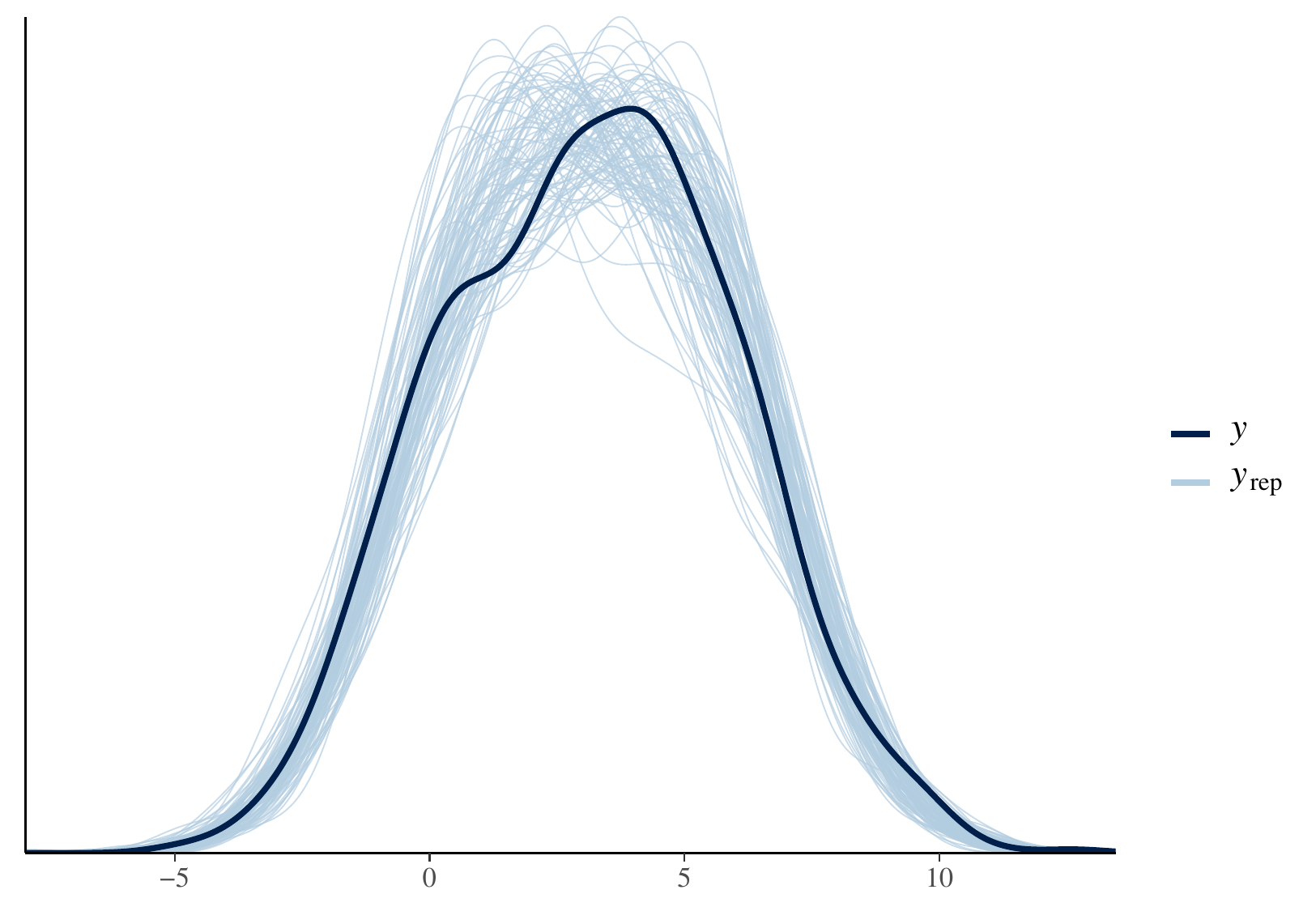} \end{center}

We also plot the autocorrelation function of the observed data and
compare to 90\% credible intervals for the ACF of the replicated data
sets.

\begin{Shaded}
\begin{Highlighting}[]
\NormalTok{nlags <-}\StringTok{ }\DecValTok{61}
\NormalTok{oac =}\StringTok{ }\KeywordTok{acf}\NormalTok{(y[}\DecValTok{2}\OperatorTok{:}\NormalTok{(n }\OperatorTok{-}\StringTok{ }\DecValTok{1}\NormalTok{)], }\DataTypeTok{lag.max =}\NormalTok{ (nlags }\OperatorTok{-}\StringTok{ }\DecValTok{1}\NormalTok{), }\DataTypeTok{plot =} \OtherTok{FALSE}\NormalTok{)}\OperatorTok{$}\NormalTok{acf  }\CommentTok{#  observed acf}

\NormalTok{ppac =}\StringTok{ }\KeywordTok{matrix}\NormalTok{(}\OtherTok{NA}\NormalTok{, n.sims, nlags)}
\ControlFlowTok{for}\NormalTok{ (i }\ControlFlowTok{in} \DecValTok{1}\OperatorTok{:}\NormalTok{n.sims) \{}
\NormalTok{    ppac[i, ] =}\StringTok{ }\KeywordTok{acf}\NormalTok{(ppobs[i, ], }\DataTypeTok{lag.max =}\NormalTok{ (nlags }\OperatorTok{-}\StringTok{ }\DecValTok{1}\NormalTok{), }\DataTypeTok{plot =} \OtherTok{FALSE}\NormalTok{)}\OperatorTok{$}\NormalTok{acf}
\NormalTok{\}}

\NormalTok{hpd.acf <-}\StringTok{ }\KeywordTok{HPDinterval}\NormalTok{(}\KeywordTok{as.mcmc}\NormalTok{(ppac), }\DataTypeTok{prob =} \FloatTok{0.95}\NormalTok{)}
\NormalTok{dat <-}\StringTok{ }\KeywordTok{data.frame}\NormalTok{(}\DataTypeTok{x =} \DecValTok{1}\OperatorTok{:}\DecValTok{61}\NormalTok{, }\DataTypeTok{acf =} \KeywordTok{as.numeric}\NormalTok{(oac), }\DataTypeTok{lb =}\NormalTok{ hpd.acf[, }\DecValTok{1}\NormalTok{], }\DataTypeTok{ub =}\NormalTok{ hpd.acf[, }
    \DecValTok{2}\NormalTok{])}

\KeywordTok{ggplot}\NormalTok{(dat, }\KeywordTok{aes}\NormalTok{(x, acf)) }\OperatorTok{+}\StringTok{ }\KeywordTok{geom_ribbon}\NormalTok{(}\KeywordTok{aes}\NormalTok{(}\DataTypeTok{x =}\NormalTok{ x, }\DataTypeTok{ymin =}\NormalTok{ lb, }\DataTypeTok{ymax =}\NormalTok{ ub), }\DataTypeTok{fill =} \StringTok{"grey70"}\NormalTok{, }
    \DataTypeTok{alpha =} \FloatTok{0.5}\NormalTok{) }\OperatorTok{+}\StringTok{ }\KeywordTok{geom_point}\NormalTok{(}\DataTypeTok{col =} \StringTok{"purple"}\NormalTok{, }\DataTypeTok{size =} \DecValTok{1}\NormalTok{) }\OperatorTok{+}\StringTok{ }\KeywordTok{geom_line}\NormalTok{() }\OperatorTok{+}\StringTok{ }\KeywordTok{coord_cartesian}\NormalTok{(}\DataTypeTok{xlim =} \KeywordTok{c}\NormalTok{(}\DecValTok{2}\NormalTok{, }
    \DecValTok{60}\NormalTok{), }\DataTypeTok{ylim =} \KeywordTok{c}\NormalTok{(}\OperatorTok{-}\FloatTok{0.1}\NormalTok{, }\FloatTok{0.5}\NormalTok{)) }\OperatorTok{+}\StringTok{ }\KeywordTok{xlab}\NormalTok{(}\StringTok{"Lag"}\NormalTok{) }\OperatorTok{+}\StringTok{ }\KeywordTok{ylab}\NormalTok{(}\StringTok{"ACF"}\NormalTok{) }\OperatorTok{+}\StringTok{ }\KeywordTok{ggtitle}\NormalTok{(}\StringTok{"Observed Autocorrelation }
\StringTok{  Function with 90% CI for ACF of Predicted Quantities"}\NormalTok{)}
\end{Highlighting}
\end{Shaded}

\begin{center}\includegraphics[width=0.7\linewidth]{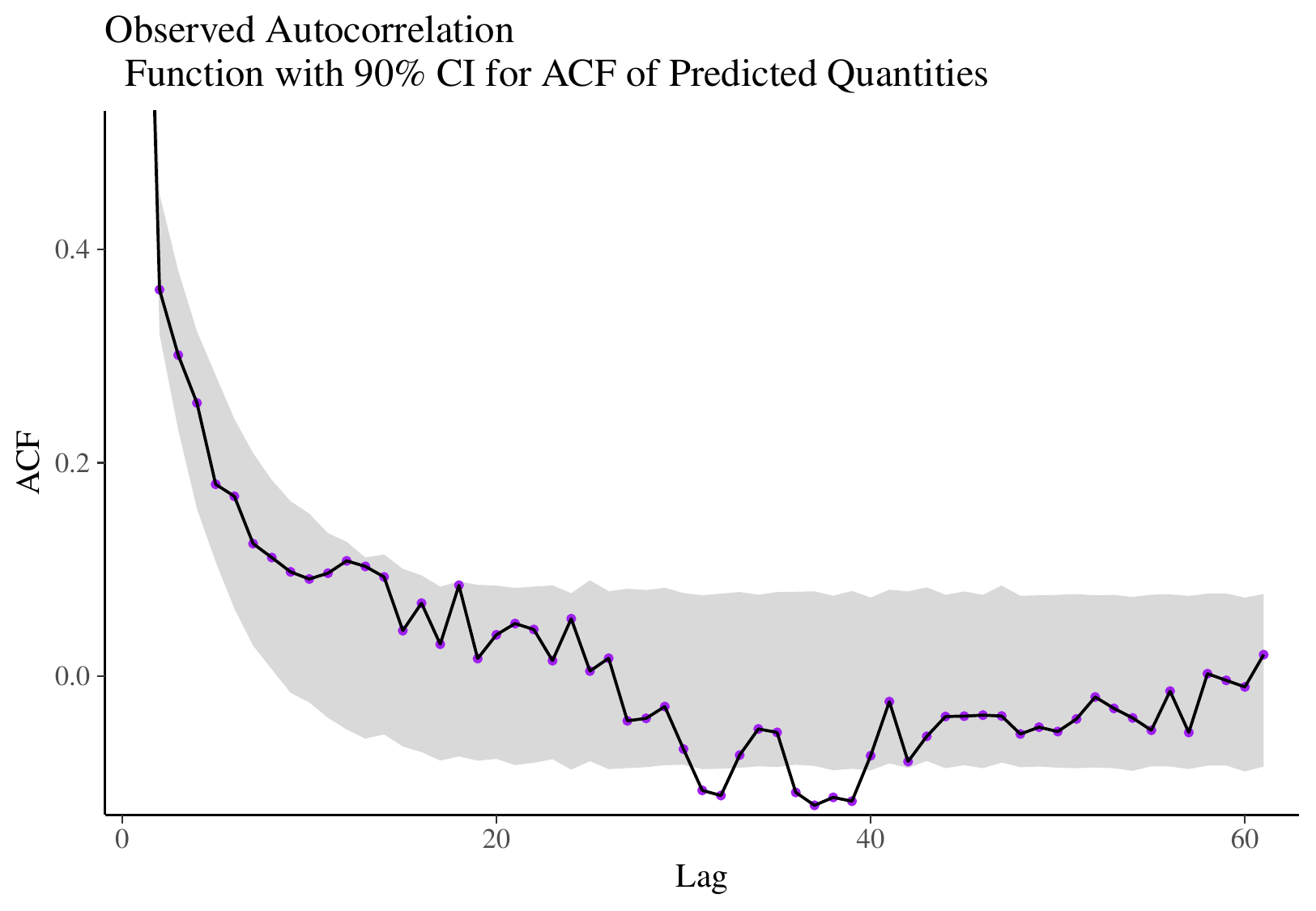} \end{center}

Finally, we use the posterior expected values of the variables of
interest to construct the forecast (pseudo-)residuals.

\begin{Shaded}
\begin{Highlighting}[]
\NormalTok{##### R code for Forecast (Pseudo-)Residuals}

\NormalTok{## Calculating forward variables}
\NormalTok{HMM.lalpha <-}\StringTok{ }\ControlFlowTok{function}\NormalTok{(allprobs, gamma, delta, n, N, mu) \{}
\NormalTok{    lalpha <-}\StringTok{ }\KeywordTok{matrix}\NormalTok{(}\OtherTok{NA}\NormalTok{, N, n)}
    
\NormalTok{    lscale <-}\StringTok{ }\DecValTok{0}
\NormalTok{    foo <-}\StringTok{ }\NormalTok{delta }\OperatorTok{*}\StringTok{ }\NormalTok{allprobs[}\DecValTok{1}\NormalTok{, ]}
\NormalTok{    lscale <-}\StringTok{ }\DecValTok{0}
\NormalTok{    lalpha[, }\DecValTok{1}\NormalTok{] <-}\StringTok{ }\KeywordTok{log}\NormalTok{(foo) }\OperatorTok{+}\StringTok{ }\NormalTok{lscale}
\NormalTok{    sumfoo <-}\StringTok{ }\KeywordTok{sum}\NormalTok{(foo)}
    \ControlFlowTok{for}\NormalTok{ (i }\ControlFlowTok{in} \DecValTok{2}\OperatorTok{:}\NormalTok{n) \{}
\NormalTok{        foo <-}\StringTok{ }\NormalTok{foo }\OperatorTok{%*%}\StringTok{ }\NormalTok{gamma }\OperatorTok{*}\StringTok{ }\NormalTok{allprobs[i, ]}
\NormalTok{        sumfoo <-}\StringTok{ }\KeywordTok{sum}\NormalTok{(foo)}
\NormalTok{        lscale <-}\StringTok{ }\NormalTok{lscale }\OperatorTok{+}\StringTok{ }\KeywordTok{log}\NormalTok{(sumfoo)}
\NormalTok{        foo <-}\StringTok{ }\NormalTok{foo}\OperatorTok{/}\NormalTok{sumfoo  }\CommentTok{# scaling}
\NormalTok{        lalpha[, i] <-}\StringTok{ }\KeywordTok{log}\NormalTok{(foo) }\OperatorTok{+}\StringTok{ }\NormalTok{lscale}
\NormalTok{    \}}
\NormalTok{    lalpha}
\NormalTok{\}}


\NormalTok{## Calculating forecast (pseudo-)residuals}
\NormalTok{HMM.psres <-}\StringTok{ }\ControlFlowTok{function}\NormalTok{(x, allprobs, gamma, n, N, mu) \{}
    
    
\NormalTok{    delta <-}\StringTok{ }\KeywordTok{solve}\NormalTok{(}\KeywordTok{t}\NormalTok{(}\KeywordTok{diag}\NormalTok{(N) }\OperatorTok{-}\StringTok{ }\NormalTok{gamma }\OperatorTok{+}\StringTok{ }\DecValTok{1}\NormalTok{), }\KeywordTok{rep}\NormalTok{(}\DecValTok{1}\NormalTok{, N))}
    
\NormalTok{    la <-}\StringTok{ }\KeywordTok{HMM.lalpha}\NormalTok{(allprobs, gamma, delta, n, N, mu)}
    
\NormalTok{    pstepmat <-}\StringTok{ }\KeywordTok{matrix}\NormalTok{(}\OtherTok{NA}\NormalTok{, n, N)}
\NormalTok{    fres <-}\StringTok{ }\KeywordTok{rep}\NormalTok{(}\OtherTok{NA}\NormalTok{, n)}
\NormalTok{    ind.step <-}\StringTok{ }\KeywordTok{which}\NormalTok{(}\OperatorTok{!}\KeywordTok{is.na}\NormalTok{(x))}
    
    \ControlFlowTok{for}\NormalTok{ (j }\ControlFlowTok{in} \DecValTok{1}\OperatorTok{:}\KeywordTok{length}\NormalTok{(ind.step)) \{}
\NormalTok{        pstepmat[ind.step[j], }\DecValTok{1}\NormalTok{] <-}\StringTok{ }\KeywordTok{pnorm}\NormalTok{(x[ind.step[j]], }\DataTypeTok{mean =}\NormalTok{ mu[}\DecValTok{1}\NormalTok{], }\DataTypeTok{sd =} \DecValTok{2}\NormalTok{)}
\NormalTok{        pstepmat[ind.step[j], }\DecValTok{2}\NormalTok{] <-}\StringTok{ }\KeywordTok{pnorm}\NormalTok{(x[ind.step[j]], }\DataTypeTok{mean =}\NormalTok{ mu[}\DecValTok{2}\NormalTok{], }\DataTypeTok{sd =} \DecValTok{2}\NormalTok{)}
\NormalTok{    \}}
    
    
    \ControlFlowTok{if}\NormalTok{ (}\OperatorTok{!}\KeywordTok{is.na}\NormalTok{(x[}\DecValTok{1}\NormalTok{])) }
\NormalTok{        fres[}\DecValTok{1}\NormalTok{] <-}\StringTok{ }\KeywordTok{qnorm}\NormalTok{(}\KeywordTok{rbind}\NormalTok{(}\KeywordTok{c}\NormalTok{(}\DecValTok{1}\NormalTok{, }\DecValTok{0}\NormalTok{)) }\OperatorTok{%*%}\StringTok{ }\NormalTok{pstepmat[}\DecValTok{1}\NormalTok{, ])}
    \ControlFlowTok{for}\NormalTok{ (i }\ControlFlowTok{in} \DecValTok{2}\OperatorTok{:}\NormalTok{n) \{}
        
\NormalTok{        c <-}\StringTok{ }\KeywordTok{max}\NormalTok{(la[, i }\OperatorTok{-}\StringTok{ }\DecValTok{1}\NormalTok{])}
\NormalTok{        a <-}\StringTok{ }\KeywordTok{exp}\NormalTok{(la[, i }\OperatorTok{-}\StringTok{ }\DecValTok{1}\NormalTok{] }\OperatorTok{-}\StringTok{ }\NormalTok{c)}
        \ControlFlowTok{if}\NormalTok{ (}\OperatorTok{!}\KeywordTok{is.na}\NormalTok{(x[i])) }
\NormalTok{            fres[i] <-}\StringTok{ }\KeywordTok{qnorm}\NormalTok{(}\KeywordTok{t}\NormalTok{(a) }\OperatorTok{%*%}\StringTok{ }\NormalTok{(gamma}\OperatorTok{/}\KeywordTok{sum}\NormalTok{(a)) }\OperatorTok{%*%}\StringTok{ }\NormalTok{pstepmat[i, ])}
\NormalTok{    \}}
    \KeywordTok{return}\NormalTok{(}\KeywordTok{list}\NormalTok{(}\DataTypeTok{fres =}\NormalTok{ fres))}
\NormalTok{\}}

\NormalTok{means <-}\StringTok{ }\KeywordTok{colMeans}\NormalTok{(mus[[}\DecValTok{1}\NormalTok{]])}

\NormalTok{allprobs <-}\StringTok{ }\KeywordTok{matrix}\NormalTok{(}\DecValTok{1}\NormalTok{, }\DataTypeTok{nrow =}\NormalTok{ n, }\DataTypeTok{ncol =}\NormalTok{ N)}
\ControlFlowTok{for}\NormalTok{ (j }\ControlFlowTok{in} \DecValTok{1}\OperatorTok{:}\NormalTok{N) allprobs[}\KeywordTok{which}\NormalTok{(}\OperatorTok{!}\KeywordTok{is.na}\NormalTok{(y)), j] <-}\StringTok{ }\KeywordTok{dnorm}\NormalTok{(y, }\DataTypeTok{mean =}\NormalTok{ means[j], }\DataTypeTok{sd =} \DecValTok{2}\NormalTok{)}

\NormalTok{gamma <-}\StringTok{ }\KeywordTok{matrix}\NormalTok{(}\KeywordTok{c}\NormalTok{(}\KeywordTok{mean}\NormalTok{(psam[[}\DecValTok{1}\NormalTok{]][, }\DecValTok{1}\NormalTok{, }\DecValTok{1}\NormalTok{]), }\DecValTok{1} \OperatorTok{-}\StringTok{ }\KeywordTok{mean}\NormalTok{(psam[[}\DecValTok{1}\NormalTok{]][, }\DecValTok{1}\NormalTok{, }\DecValTok{1}\NormalTok{]), }\DecValTok{1} \OperatorTok{-}\StringTok{ }
\StringTok{    }\KeywordTok{mean}\NormalTok{(psam[[}\DecValTok{1}\NormalTok{]][, }\DecValTok{2}\NormalTok{, }\DecValTok{2}\NormalTok{]), }\KeywordTok{mean}\NormalTok{(psam[[}\DecValTok{1}\NormalTok{]][, }\DecValTok{2}\NormalTok{, }\DecValTok{2}\NormalTok{])), }\DataTypeTok{nrow =} \DecValTok{2}\NormalTok{, }\DataTypeTok{byrow =}\NormalTok{ T)}

\NormalTok{fres <-}\StringTok{ }\KeywordTok{HMM.psres}\NormalTok{(}\DataTypeTok{x =}\NormalTok{ y, }\DataTypeTok{allprobs =}\NormalTok{ allprobs, }\DataTypeTok{gamma =}\NormalTok{ gamma, }\DataTypeTok{n =}\NormalTok{ n, }\DataTypeTok{N =}\NormalTok{ N, }\DataTypeTok{mu =}\NormalTok{ means)}
\end{Highlighting}
\end{Shaded}

Plotting the residuals in a Q-Q plot:

\begin{Shaded}
\begin{Highlighting}[]
\KeywordTok{ggplot}\NormalTok{(}\DataTypeTok{data=}\KeywordTok{data.frame}\NormalTok{(}\DataTypeTok{x=}\NormalTok{fres}\OperatorTok{$}\NormalTok{fres), }\KeywordTok{aes}\NormalTok{(}\DataTypeTok{sample =}\NormalTok{ x)) }\OperatorTok{+}\StringTok{ }\KeywordTok{stat_qq}\NormalTok{() }\OperatorTok{+}\StringTok{ }
\StringTok{  }\KeywordTok{stat_qq_line}\NormalTok{(}\DataTypeTok{color=}\StringTok{"purple"}\NormalTok{, }\DataTypeTok{size=}\DecValTok{1}\NormalTok{) }\OperatorTok{+}
\StringTok{  }\KeywordTok{ggtitle}\NormalTok{(}\StringTok{"Q-Q Plot"}\NormalTok{) }\OperatorTok{+}\StringTok{ }\KeywordTok{theme_classic}\NormalTok{()}
\end{Highlighting}
\end{Shaded}

\begin{center}\includegraphics[width=0.7\linewidth]{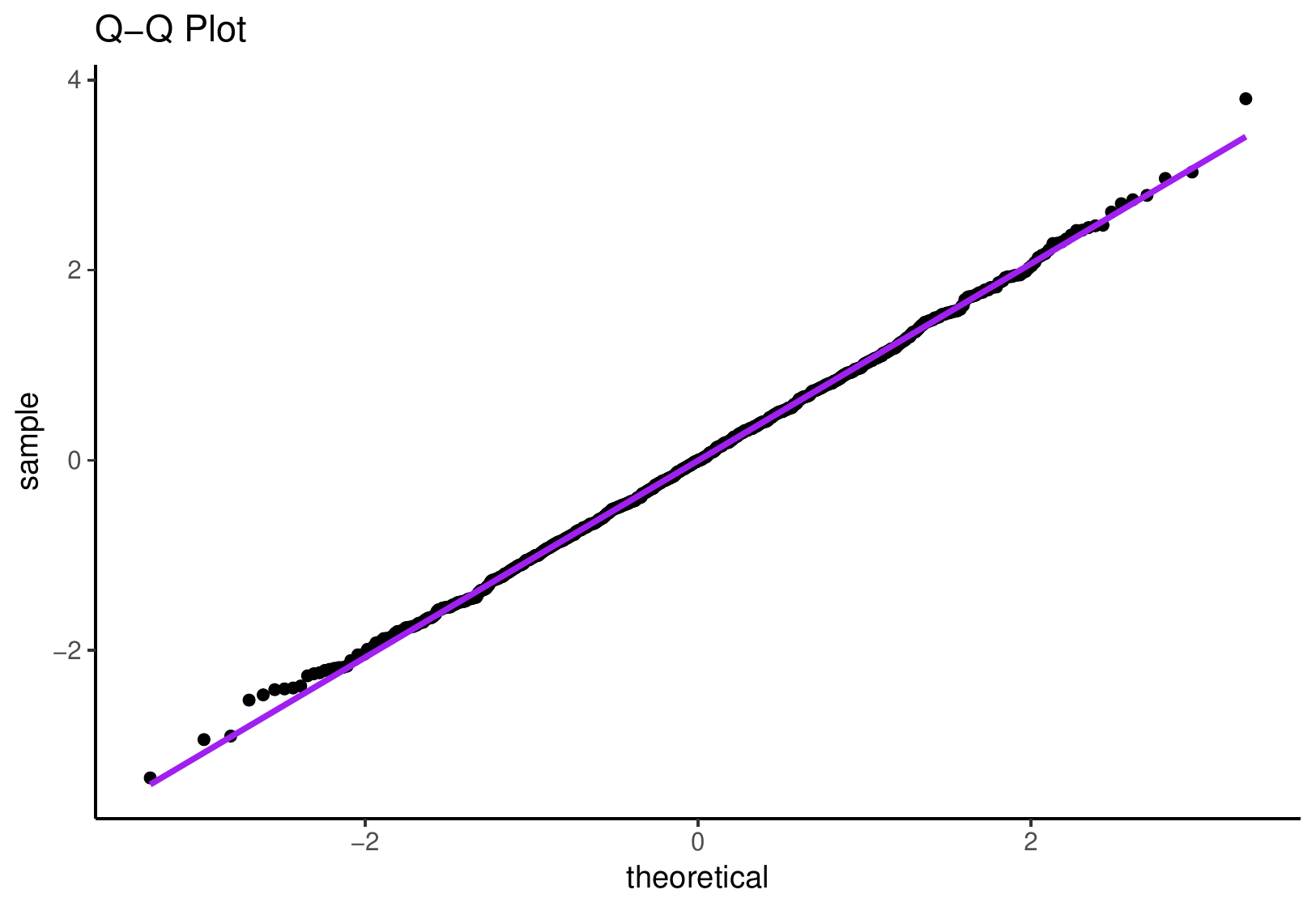} \end{center}

Note that it is also possible to construct the distribution of forecast
residuals at each time \(t\).

\section{Covariates}\label{covariates}

In HMMs applied to animal movement, covariates are typically
incorporated at the level of the hidden states. For the general case of
time-varying covariates, we define the corresponding time-dependent
transition probability matrix
\(\boldsymbol\Gamma^{(t)} = (\gamma_{ij}^{(t)})\), where
\(\gamma_{ij}^{(t)} = \Pr(S_{t+1}=j \vert S_t=i)\). The transition
probabilities at time \(t\), \(\gamma_{ij}^{(t)}\), can then be related
to a vector of environmental (or other) covariates,
\(\bigl(\omega_1^{(t)},\dots,\omega_p^{(t)}\bigr)\), via the multinomial
logit link:

\begin{equation*}
\gamma_{ij}^{(t)} = \dfrac{\exp(\eta_{ij})}{\sum_{k=1}^N \exp(\eta_{ik})}, \qquad \text{where} \qquad
\eta_{ij} = 
\begin{cases}
\beta_0^{(ij)} + \sum_{l=1}^p \beta_l^{(ij)} \omega_l^{(t)} & \text{ if } i \neq j;\\
0 & \text{ otherwise}.
\end{cases}
\end{equation*}

Essentially there is one multinomial logit link specification for each
row of the matrix \(\boldsymbol\Gamma^{(t)}\), and the entries on the
diagonal of the matrix serve as reference categories.

\section{Modeling Animal Movement with
HMMs}\label{modeling-animal-movement-with-hmms}

\subsection{Motivation}\label{motivation}

We consider the application of HMMs to the analysis of animal movement
tracks. Movement data typically consist of a bivariate time series of
longitude-latitude positions, collected at regular time intervals over
the study period (e.g.~hourly locations). HMMs are widely used in
movement ecology to describe such data as arising from several distinct
movement patterns, modelled by the underlying Markov chain \(S_t\). In
particular, these movement patterns serve as proxies for general
behaviors of interest. At each time step, we consider that an animal is
in one of \(N\) (behavioural) states (e.g. ``exploratory'',
``foraging''\ldots{}), on which depend some metrics of movement. Note
that there is generally no 1-1 mapping from state to behavior of
interest, but more on this later.

In this context, the most common HMM formulation is based on the step
lengths and turning angles, which can be derived from the location data.
The step length \(L_t\) is the distance between the two successive
locations \(X_t\) and \(X_{t+1}\), and the turning angle \(\varphi_t\)
is the angle between the two successive directions \((X_{t-1},X_t)\) and
\((X_t,X_{t+1})\).

\subsection{Wild Haggis}\label{wild-haggis}

We present a simulation study based on the (simulated) wild haggis
tracking data from Michelot et al. (2016). The data set comprises 15
tracks, with slope and temperature covariates.

\begin{figure}[htbp]
\centering
\includegraphics[width=0.3\textwidth]{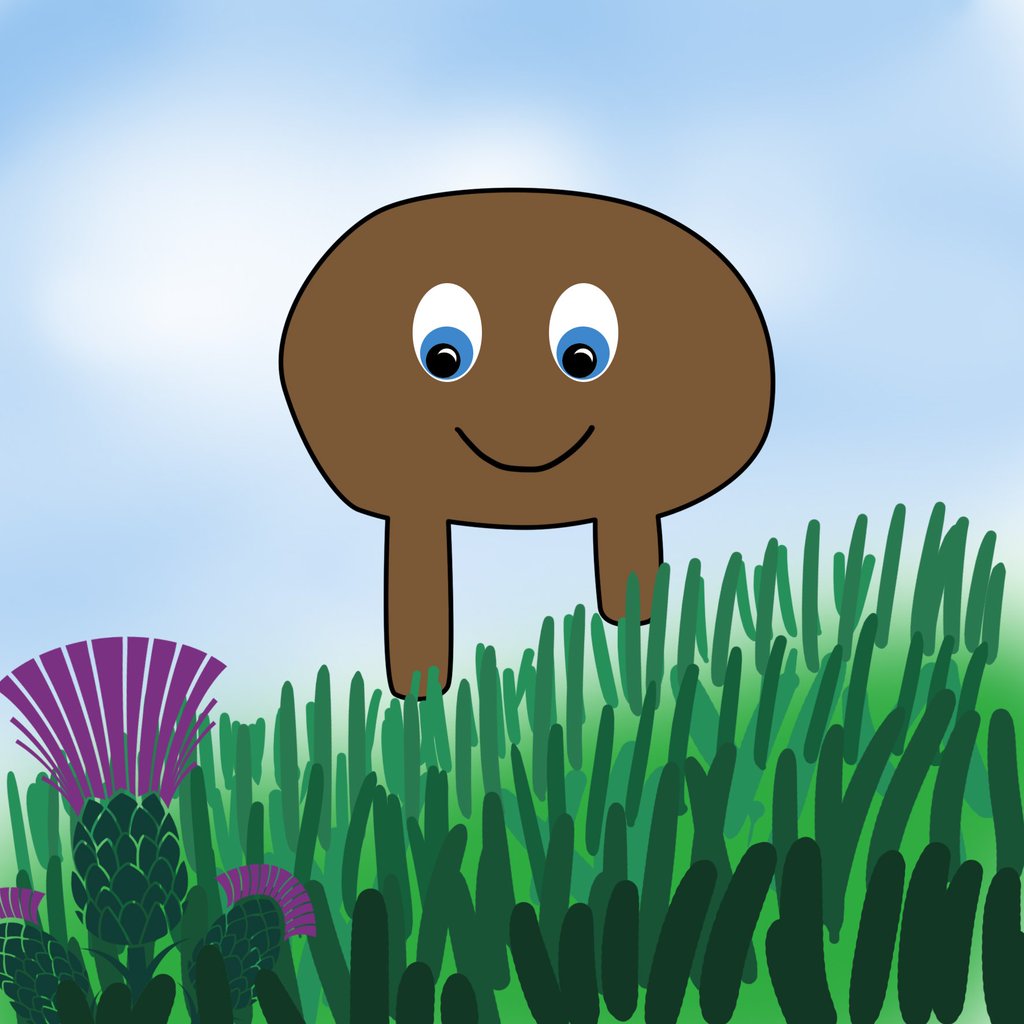}
\caption{Andrea Langrock's impression of the elusive wild haggis}
\end{figure}

We use the function \texttt{prepData} in the package moveHMM to derive
step lengths and turning angles from the location data.

\begin{Shaded}
\begin{Highlighting}[]
\NormalTok{rawhaggis <-}\StringTok{ }\KeywordTok{read.csv}\NormalTok{(}\StringTok{"data/haggis.csv"}\NormalTok{)}
\CommentTok{# derive step lengths and turning angles from locations}
\NormalTok{data <-}\StringTok{ }\KeywordTok{prepData}\NormalTok{(rawhaggis, }\DataTypeTok{type=}\StringTok{"UTM"}\NormalTok{)}

\KeywordTok{hist}\NormalTok{(data}\OperatorTok{$}\NormalTok{step, }\DataTypeTok{main=}\StringTok{""}\NormalTok{, }\DataTypeTok{xlab=}\StringTok{"Step length"}\NormalTok{)}
\KeywordTok{hist}\NormalTok{(data}\OperatorTok{$}\NormalTok{angle, }\DataTypeTok{breaks=}\KeywordTok{seq}\NormalTok{(}\OperatorTok{-}\NormalTok{pi,pi,}\DataTypeTok{length=}\DecValTok{15}\NormalTok{), }\DataTypeTok{main=}\StringTok{""}\NormalTok{, }\DataTypeTok{xlab=}\StringTok{"Turning angle"}\NormalTok{)}
\end{Highlighting}
\end{Shaded}

\begin{figure}
\includegraphics[width=0.49\linewidth]{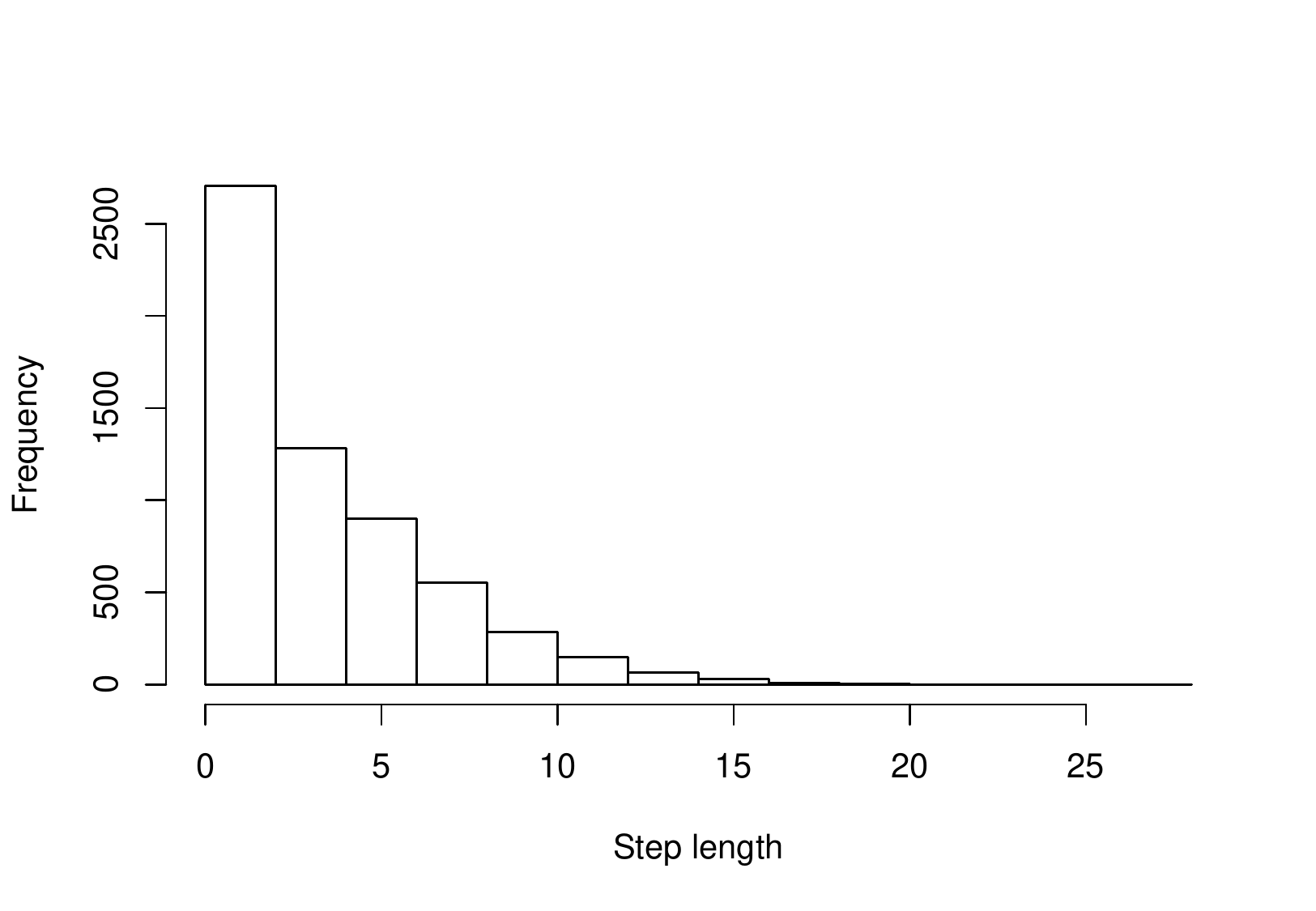} \includegraphics[width=0.49\linewidth]{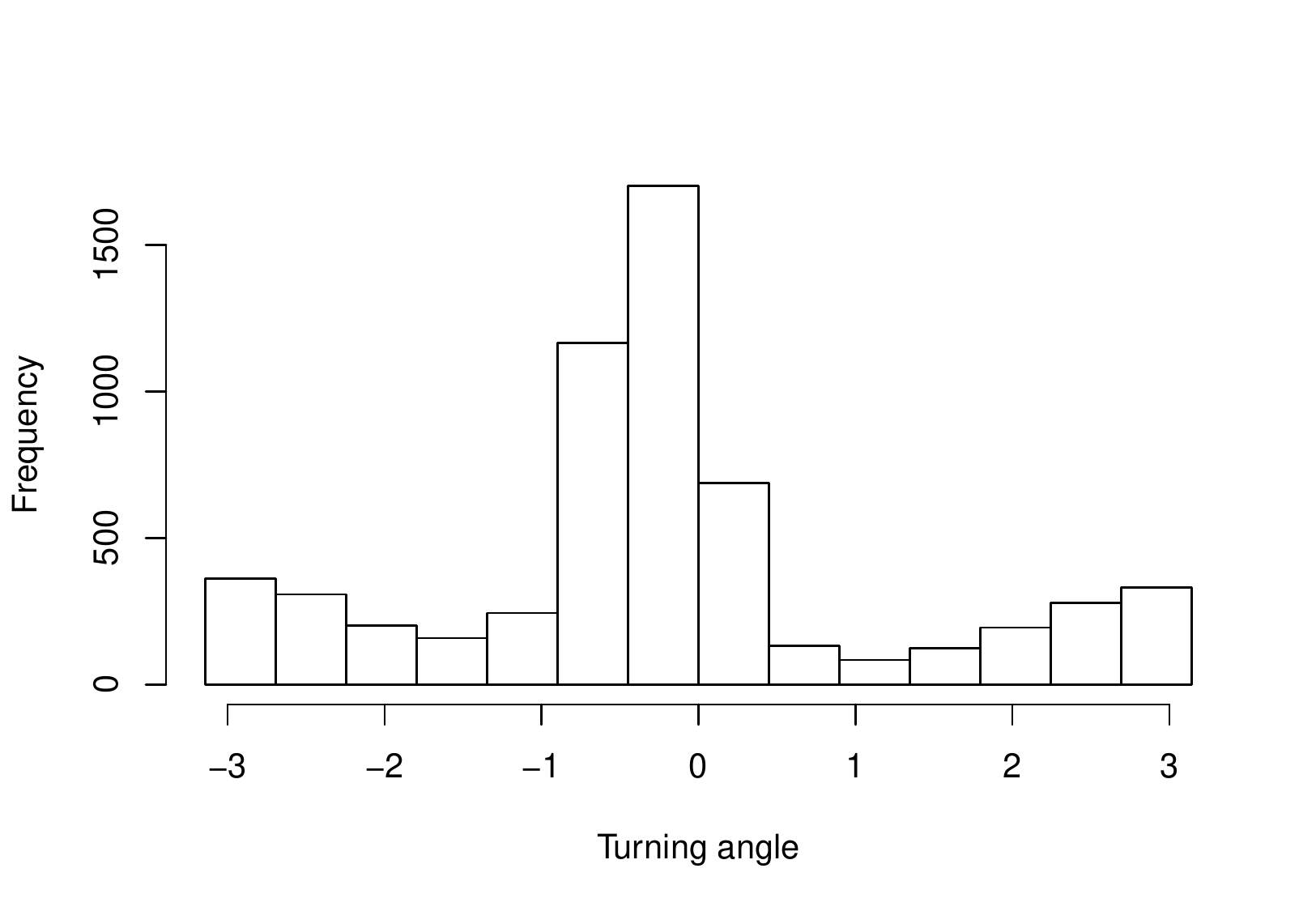} \caption{Histograms of the step lengths (left) and turning angles (right) in the wild haggis data.}\label{fig:unnamed-chunk-15}
\end{figure}

Following Michelot et al. (2016), we consider a 2-state HMM with gamma
and von Mises state-dependent distributions. That is, for
\(j \in \{ 1,2 \}\)

\begin{align*}
L_t \vert S_t=j & \sim \text{gamma}(\alpha_j, \beta_j) \\
\varphi_t \vert S_t=j & \sim \text{von Mises}(\mu_j, \kappa_j),
\end{align*}

where \(\alpha_j\) is the shape and \(\beta_j\) the rate of the gamma
distribution, and \(\mu_j\) is the mean and \(\kappa_j\) the
concentration of the von Mises distribution. The larger the
concentration, the smaller the variance of the turning angles around
their mean.

We find it more convenient to parametrise the gamma distribution in
terms of its mean and standard deviation, rather than its scale and rate
parameters (default in R and Stan). We use the following transformation
to obtain one set of parameters from the other:

\begin{equation*}
\text{shape} = \dfrac{\text{mean}^2}{\text{SD}^2},
\quad \text{rate} = \dfrac{\text{mean}}{\text{SD}^2}.
\end{equation*}

The mean parameter of the von Mises distribution is constrained between
\(-\pi\) and \(\pi\). This can cause estimation issues, if the sampler
gets stuck around either bound. To address this problem, we consider the
alternative parametrisation: for each state \(j\),

\begin{equation}
\label{eqn:anglevars}
\begin{cases}
x^\varphi_j = \kappa_j \cos (\mu_j)\\
y^\varphi_j = \kappa_j \sin (\mu_j)
\end{cases}
\end{equation}

The point \((x^\varphi_j,y^\varphi_j)\) is unconstrained in
\(\mathbb{R}^2\).

The following code implements a \(N\)-state HMM with gamma and von Mises
state-dependent distributions, with the possibility to include
covariates in the state process. We describe each block separately.

\begin{verbatim}
data {
    int<lower=0> T; // length of the time series
    int ID[T]; // track identifier
    vector[T] steps; // step lengths
    vector[T] angles; // turning angles
    int<lower=1> N; // number of states
    int nCovs; // number of covariates
    matrix[T,nCovs+1] covs; // covariates
}
\end{verbatim}

In the `data' block, we include the vector of step lengths, the vector
of turning angles, and the (design) matrix of covariate values. The
design matrix has one column of 1s, corresponding to the intercept, and
one column for each covariate. We also need to specify the length of the
time series (i.e.~number of locations), the number of states (two in the
analysis), and the number of covariates (three in the analysis:
temperature, slope, and slope\(^2\)).

\begin{verbatim}
parameters {
    positive_ordered[N] mu; // mean of gamma - ordered
    vector<lower=0>[N] sigma; // SD of gamma
    // unconstrained angle parameters
    vector[N] xangle;
    vector[N] yangle;
    // regression coefficients for transition probabilities
    matrix[N*(N-1),nCovs+1] beta; 
}  
\end{verbatim}

We define the state-dependent movement parameters: the mean and standard
deviation of the gamma distribution (step lengths), and the transformed
unconstrained parameters of the turning angle distribution defined in
Equation \ref{eqn:anglevars}. The vector of mean step lengths is defined
to be ordered, to avoid label switching. We also introduce the matrix of
regression coefficients for the transition probabilities, with one row
for each non-diagonal entry of the transition probability matrix, and
one column for each covariable (plus one for the intercept).

\begin{verbatim}
transformed parameters {
    vector<lower=0>[N] shape;
    vector<lower=0>[N] rate;
    vector<lower=-pi(),upper=pi()>[N] loc;
    vector<lower=0>[N] kappa;

    
    // derive turning angle mean and concentration
    for(n in 1:N) {
        loc[n] = atan2(yangle[n], xangle[n]);
        kappa[n] = sqrt(xangle[n]*xangle[n] + yangle[n]*yangle[n]);
    }
    
    // transform mean and SD to shape and rate
    for(n in 1:N)
        shape[n] = mu[n]*mu[n]/(sigma[n]*sigma[n]);
    
    for(n in 1:N)
        rate[n] = mu[n]/(sigma[n]*sigma[n]);
}
\end{verbatim}

In the `transformed parameters', we calculate the parameters expected by
the state-dependent pdfs, i.e.~the shape and rate of the gamma
distribution, and the location (mean) and concentration of the von Mises
distribution.

\begin{verbatim}
model {
    vector[N] logp;
    vector[N] logptemp;
    matrix[N,N] gamma[T];
    matrix[N,N] log_gamma[T];
    matrix[N,N] log_gamma_tr[T];
    
    // priors
    mu ~ normal(0, 5);
    sigma ~ student_t(3, 0, 1);
    xangle[1] ~ normal(-0.5, 1); // equiv to concentration when yangle = 0
    xangle[2] ~ normal(2, 2);
    yangle ~ normal(0, 0.5); // zero if mean angle is 0 or pi
      
    // derive array of (log-)transition probabilities
    for(t in 1:T) {
        int betarow = 1;
        for(i in 1:N) {
            for(j in 1:N) {
                if(i==j) {
                    gamma[t,i,j] = 1;
                } else {
                    gamma[t,i,j] = exp(beta[betarow] * to_vector(covs[t]));
                    betarow = betarow + 1;
                }
            }
        }
        
        // each row must sum to 1
        for(i in 1:N)
            log_gamma[t][i] = log(gamma[t][i]/sum(gamma[t][i]));
    }
    
    // transpose
    for(t in 1:T)
        for(i in 1:N)
            for(j in 1:N)
                log_gamma_tr[t,j,i] = log_gamma[t,i,j];

    // likelihood computation
    for (t in 1:T) {
        // initialise forward variable if first obs of track
        if(t==1 || ID[t]!=ID[t-1])
            logp = rep_vector(-log(N), N);
        
        for (n in 1:N) {
            logptemp[n] = log_sum_exp(to_vector(log_gamma_tr[t,n]) + logp);
            if(steps[t]>=0)
                logptemp[n] = logptemp[n] + gamma_lpdf(steps[t] | shape[n], rate[n]);
            if(angles[t]>=(-pi()))
                logptemp[n] = logptemp[n] + von_mises_lpdf(angles[t] | loc[n], kappa[n]);    
        }
        logp = logptemp;
        
        // add log forward variable to target at the end of each track
        if(t==T || ID[t+1]!=ID[t])
            target += log_sum_exp(logp);
    }
}
\end{verbatim}

We derive the transition probability matrix \(\boldsymbol\Gamma^{(t)}\),
at each time point, from the regression coefficients and the covariates
values provided. We store the log transition probabilities, which we use
in the forward algorithm, in the array \(\texttt{log\_gamma\_tr}\). Note
that each matrix (each layer of the array) is transposed, so that each
row corresponds to the probabilities of transitioning into a state,
rather than out of a state.

We choose priors on the movement parameters based on previous biological
knowledge of the movements of the wild haggis.

The loop over the observations corresponds to the forward algorithm, on
the log-scale to obtain the log-likelihood and circumvent numerical
problems. At time \(t\), the \(j\)-th element of the log forward
variable can be written as

\begin{align*}
\log(\alpha_{t,j}) & = 
\log \left( \sum_{i=1}^N \gamma_{ij} \alpha_{t-1,i} \right) \\
& = \log \left( \sum_{i=1}^N \exp ( \log(\gamma_{ij}) + \log(\alpha_{t-1,i}) ) \right)
\end{align*}

where the \(\{ \log(\gamma_{ij}) \}_{i=1}^N\) are given by the \(j\)-th
row of the (transposed) matrix of log transition probabilities
\(\texttt{log\_gamma\_tr}\), and the \(\log(\alpha_{t-1,i})\) are
obtained iteratively.

We fit the model to the haggis data.

\begin{Shaded}
\begin{Highlighting}[]
\CommentTok{# set NAs to out-of-range values}
\NormalTok{data}\OperatorTok{$}\NormalTok{step[}\KeywordTok{is.na}\NormalTok{(data}\OperatorTok{$}\NormalTok{step)] <-}\StringTok{ }\OperatorTok{-}\DecValTok{10}
\NormalTok{data}\OperatorTok{$}\NormalTok{angle[}\KeywordTok{is.na}\NormalTok{(data}\OperatorTok{$}\NormalTok{angle)] <-}\StringTok{ }\OperatorTok{-}\DecValTok{10}
\NormalTok{data}\OperatorTok{$}\NormalTok{ID <-}\StringTok{ }\KeywordTok{as.numeric}\NormalTok{(data}\OperatorTok{$}\NormalTok{ID)}

\NormalTok{stan.data <-}\StringTok{ }\KeywordTok{list}\NormalTok{(}\DataTypeTok{T=}\KeywordTok{nrow}\NormalTok{(data), }\DataTypeTok{ID=}\NormalTok{data}\OperatorTok{$}\NormalTok{ID, }\DataTypeTok{steps=}\NormalTok{data}\OperatorTok{$}\NormalTok{step, }\DataTypeTok{angles=}\NormalTok{data}\OperatorTok{$}\NormalTok{angle, }\DataTypeTok{N=}\DecValTok{2}\NormalTok{, }\DataTypeTok{nCovs=}\DecValTok{3}\NormalTok{,}
                  \DataTypeTok{covs=}\KeywordTok{cbind}\NormalTok{(}\DecValTok{1}\NormalTok{, }\KeywordTok{scale}\NormalTok{(data}\OperatorTok{$}\NormalTok{temp), }\KeywordTok{scale}\NormalTok{(data}\OperatorTok{$}\NormalTok{slope), }\KeywordTok{scale}\NormalTok{(data}\OperatorTok{$}\NormalTok{slope)}\OperatorTok{^}\DecValTok{2}\NormalTok{))}

\NormalTok{inits <-}\StringTok{ }\KeywordTok{list}\NormalTok{(}\KeywordTok{list}\NormalTok{(}\DataTypeTok{mu=}\KeywordTok{c}\NormalTok{(}\DecValTok{1}\NormalTok{,}\DecValTok{5}\NormalTok{), }\DataTypeTok{sigma=}\KeywordTok{c}\NormalTok{(}\DecValTok{1}\NormalTok{,}\DecValTok{5}\NormalTok{), }\DataTypeTok{xangle=}\KeywordTok{c}\NormalTok{(}\OperatorTok{-}\DecValTok{1}\NormalTok{,}\DecValTok{3}\NormalTok{), }
                   \DataTypeTok{yangle=}\KeywordTok{c}\NormalTok{(}\DecValTok{0}\NormalTok{,}\DecValTok{0}\NormalTok{), }\DataTypeTok{beta=}\KeywordTok{matrix}\NormalTok{(}\KeywordTok{c}\NormalTok{(}\OperatorTok{-}\DecValTok{2}\NormalTok{,}\OperatorTok{-}\DecValTok{2}\NormalTok{,}\DecValTok{0}\NormalTok{,}\DecValTok{0}\NormalTok{,}\DecValTok{0}\NormalTok{,}\DecValTok{0}\NormalTok{,}\DecValTok{0}\NormalTok{,}\DecValTok{0}\NormalTok{),}\DataTypeTok{nrow=}\DecValTok{2}\NormalTok{)),}
              \KeywordTok{list}\NormalTok{(}\DataTypeTok{mu=}\KeywordTok{c}\NormalTok{(}\DecValTok{1}\NormalTok{,}\DecValTok{5}\NormalTok{), }\DataTypeTok{sigma=}\KeywordTok{c}\NormalTok{(}\DecValTok{1}\NormalTok{,}\DecValTok{5}\NormalTok{), }\DataTypeTok{xangle=}\KeywordTok{c}\NormalTok{(}\OperatorTok{-}\DecValTok{1}\NormalTok{,}\DecValTok{3}\NormalTok{), }
                   \DataTypeTok{yangle=}\KeywordTok{c}\NormalTok{(}\DecValTok{0}\NormalTok{,}\DecValTok{0}\NormalTok{), }\DataTypeTok{beta=}\KeywordTok{matrix}\NormalTok{(}\KeywordTok{c}\NormalTok{(}\OperatorTok{-}\DecValTok{2}\NormalTok{,}\OperatorTok{-}\DecValTok{2}\NormalTok{,}\DecValTok{0}\NormalTok{,}\DecValTok{0}\NormalTok{,}\DecValTok{0}\NormalTok{,}\DecValTok{0}\NormalTok{,}\DecValTok{0}\NormalTok{,}\DecValTok{0}\NormalTok{),}\DataTypeTok{nrow=}\DecValTok{2}\NormalTok{)))}

\NormalTok{fit <-}\StringTok{ }\KeywordTok{stan}\NormalTok{(}\DataTypeTok{file=}\StringTok{"HMMmovement.stan"}\NormalTok{, }\DataTypeTok{data=}\NormalTok{stan.data, }\DataTypeTok{iter=}\DecValTok{1000}\NormalTok{, }\DataTypeTok{init=}\NormalTok{inits,}
            \DataTypeTok{control=}\KeywordTok{list}\NormalTok{(}\DataTypeTok{adapt_delta=}\FloatTok{0.9}\NormalTok{), }\DataTypeTok{chains=}\DecValTok{2}\NormalTok{)}
\end{Highlighting}
\end{Shaded}

We can obtain summaries and diagnostics from the fitted model object:

\begin{Shaded}
\begin{Highlighting}[]
\KeywordTok{get_elapsed_time}\NormalTok{(fit)}
\end{Highlighting}
\end{Shaded}

\begin{verbatim}
##          warmup  sample
## chain:1 963.890 2480.15
## chain:2 945.301 2494.70
\end{verbatim}

\begin{Shaded}
\begin{Highlighting}[]
\KeywordTok{summary}\NormalTok{(fit, }\DataTypeTok{pars =} \KeywordTok{c}\NormalTok{(}\StringTok{"shape"}\NormalTok{, }\StringTok{"rate"}\NormalTok{, }\StringTok{"loc"}\NormalTok{, }\StringTok{"kappa"}\NormalTok{), }\DataTypeTok{probs =} \KeywordTok{c}\NormalTok{(}\FloatTok{0.05}\NormalTok{, }\FloatTok{0.95}\NormalTok{))}\OperatorTok{$}\NormalTok{summary}
\end{Highlighting}
\end{Shaded}

\begin{verbatim}
##                mean      se_mean          sd         5%        95%
## shape[1]  4.1182991 0.0037457471 0.118450922  3.9278760  4.3177401
## shape[2]  2.7859819 0.0020680500 0.065397484  2.6807733  2.8941504
## rate[1]   4.1433578 0.0040786320 0.128977668  3.9304840  4.3526516
## rate[2]   0.5576037 0.0004364508 0.013801786  0.5349910  0.5814677
## loc[1]   -2.4851165 0.0596924707 1.851187485 -3.1345873  3.1300015
## loc[2]   -0.3091046 0.0001920197 0.006072196 -0.3191760 -0.2989181
## kappa[1]  1.0116248 0.0011559750 0.036555138  0.9522811  1.0730602
## kappa[2]  8.0071543 0.0063207519 0.199879724  7.6993283  8.3305498
##              n_eff      Rhat
## shape[1] 1000.0000 1.0029196
## shape[2] 1000.0000 0.9989344
## rate[1]  1000.0000 1.0029198
## rate[2]  1000.0000 0.9991912
## loc[1]    961.7489 0.9984171
## loc[2]   1000.0000 0.9989916
## kappa[1] 1000.0000 0.9988296
## kappa[2] 1000.0000 0.9988326
\end{verbatim}

We plot the estimated step length and turning angle densities for each
state.

\begin{Shaded}
\begin{Highlighting}[]
\CommentTok{# restore NAs}
\NormalTok{data}\OperatorTok{$}\NormalTok{step[data}\OperatorTok{$}\NormalTok{step }\OperatorTok{<}\StringTok{ }\DecValTok{0}\NormalTok{] <-}\StringTok{ }\OtherTok{NA}
\NormalTok{data}\OperatorTok{$}\NormalTok{angle[data}\OperatorTok{$}\NormalTok{angle }\OperatorTok{<}\StringTok{ }\NormalTok{(}\OperatorTok{-}\NormalTok{pi)] <-}\StringTok{ }\OtherTok{NA}

\CommentTok{# unpack posterior draws}
\NormalTok{shape <-}\StringTok{ }\KeywordTok{extract}\NormalTok{(fit, }\DataTypeTok{pars =} \StringTok{"shape"}\NormalTok{)}\OperatorTok{$}\NormalTok{shape}
\NormalTok{rate <-}\StringTok{ }\KeywordTok{extract}\NormalTok{(fit, }\DataTypeTok{pars =} \StringTok{"rate"}\NormalTok{)}\OperatorTok{$}\NormalTok{rate}
\NormalTok{loc <-}\StringTok{ }\KeywordTok{extract}\NormalTok{(fit, }\DataTypeTok{pars =} \StringTok{"loc"}\NormalTok{)}\OperatorTok{$}\NormalTok{loc}
\NormalTok{kappa <-}\StringTok{ }\KeywordTok{extract}\NormalTok{(fit, }\DataTypeTok{pars =} \StringTok{"kappa"}\NormalTok{)}\OperatorTok{$}\NormalTok{kappa}

\CommentTok{# indices of posterior draws to plot (thinned for visualisation purposes)}
\NormalTok{ind <-}\StringTok{ }\KeywordTok{seq}\NormalTok{(}\DecValTok{1}\NormalTok{, }\KeywordTok{nrow}\NormalTok{(shape), }\DataTypeTok{by =} \DecValTok{5}\NormalTok{)}

\CommentTok{# plot step length densities}
\NormalTok{stepgrid <-}\StringTok{ }\KeywordTok{seq}\NormalTok{(}\KeywordTok{min}\NormalTok{(data}\OperatorTok{$}\NormalTok{step, }\DataTypeTok{na.rm =} \OtherTok{TRUE}\NormalTok{), }\KeywordTok{max}\NormalTok{(data}\OperatorTok{$}\NormalTok{step, }\DataTypeTok{na.rm =} \OtherTok{TRUE}\NormalTok{), }
    \DataTypeTok{length =} \DecValTok{100}\NormalTok{)}
\KeywordTok{plot}\NormalTok{(}\OtherTok{NA}\NormalTok{, }\DataTypeTok{xlim =} \KeywordTok{c}\NormalTok{(}\DecValTok{0}\NormalTok{, }\DecValTok{20}\NormalTok{), }\DataTypeTok{ylim =} \KeywordTok{c}\NormalTok{(}\DecValTok{0}\NormalTok{, }\FloatTok{1.1}\NormalTok{), }\DataTypeTok{xlab =} \StringTok{"step length"}\NormalTok{, }\DataTypeTok{ylab =} \StringTok{"density"}\NormalTok{)}
\ControlFlowTok{for}\NormalTok{ (i }\ControlFlowTok{in}\NormalTok{ ind) \{}
    \CommentTok{# plot density for each state}
    \KeywordTok{points}\NormalTok{(stepgrid, }\KeywordTok{dgamma}\NormalTok{(stepgrid, }\DataTypeTok{shape =}\NormalTok{ shape[i, }\DecValTok{1}\NormalTok{], }\DataTypeTok{rate =}\NormalTok{ rate[i, }\DecValTok{1}\NormalTok{]), }
        \DataTypeTok{type =} \StringTok{"l"}\NormalTok{, }\DataTypeTok{lwd =} \FloatTok{0.2}\NormalTok{, }\DataTypeTok{col =} \KeywordTok{adjustcolor}\NormalTok{(pal[}\DecValTok{1}\NormalTok{], }\DataTypeTok{alpha.f =} \FloatTok{0.1}\NormalTok{))}
    \KeywordTok{points}\NormalTok{(stepgrid, }\KeywordTok{dgamma}\NormalTok{(stepgrid, }\DataTypeTok{shape =}\NormalTok{ shape[i, }\DecValTok{2}\NormalTok{], }\DataTypeTok{rate =}\NormalTok{ rate[i, }\DecValTok{2}\NormalTok{]), }
        \DataTypeTok{type =} \StringTok{"l"}\NormalTok{, }\DataTypeTok{lwd =} \FloatTok{0.2}\NormalTok{, }\DataTypeTok{col =} \KeywordTok{adjustcolor}\NormalTok{(pal[}\DecValTok{2}\NormalTok{], }\DataTypeTok{alpha.f =} \FloatTok{0.1}\NormalTok{))}
\NormalTok{\}}

\CommentTok{# plot turning angle densities}
\NormalTok{anglegrid <-}\StringTok{ }\KeywordTok{seq}\NormalTok{(}\OperatorTok{-}\NormalTok{pi, pi, }\DataTypeTok{length =} \DecValTok{100}\NormalTok{)}
\KeywordTok{plot}\NormalTok{(}\OtherTok{NA}\NormalTok{, }\DataTypeTok{xlim =} \KeywordTok{c}\NormalTok{(}\OperatorTok{-}\NormalTok{pi, pi), }\DataTypeTok{ylim =} \KeywordTok{c}\NormalTok{(}\DecValTok{0}\NormalTok{, }\FloatTok{1.2}\NormalTok{), }\DataTypeTok{xlab =} \StringTok{"turnging angle"}\NormalTok{, }\DataTypeTok{ylab =} \StringTok{"density"}\NormalTok{)}
\ControlFlowTok{for}\NormalTok{ (i }\ControlFlowTok{in}\NormalTok{ ind[}\OperatorTok{-}\DecValTok{1}\NormalTok{]) \{}
    \CommentTok{# plot density for each state}
    \KeywordTok{points}\NormalTok{(anglegrid, }\KeywordTok{dvm}\NormalTok{(anglegrid, }\DataTypeTok{mu =}\NormalTok{ loc[i, }\DecValTok{1}\NormalTok{], }\DataTypeTok{kappa =}\NormalTok{ kappa[i, }\DecValTok{1}\NormalTok{]), }\DataTypeTok{type =} \StringTok{"l"}\NormalTok{, }
        \DataTypeTok{lwd =} \FloatTok{0.2}\NormalTok{, }\DataTypeTok{col =} \KeywordTok{adjustcolor}\NormalTok{(pal[}\DecValTok{1}\NormalTok{], }\DataTypeTok{alpha.f =} \FloatTok{0.1}\NormalTok{))}
    \KeywordTok{points}\NormalTok{(anglegrid, }\KeywordTok{dvm}\NormalTok{(anglegrid, }\DataTypeTok{mu =}\NormalTok{ loc[i, }\DecValTok{2}\NormalTok{], }\DataTypeTok{kappa =}\NormalTok{ kappa[i, }\DecValTok{2}\NormalTok{]), }\DataTypeTok{type =} \StringTok{"l"}\NormalTok{, }
        \DataTypeTok{lwd =} \FloatTok{0.2}\NormalTok{, }\DataTypeTok{col =} \KeywordTok{adjustcolor}\NormalTok{(pal[}\DecValTok{2}\NormalTok{], }\DataTypeTok{alpha.f =} \FloatTok{0.1}\NormalTok{))}
\NormalTok{\}}
\end{Highlighting}
\end{Shaded}

\begin{figure}

{\centering \includegraphics[width=0.49\linewidth]{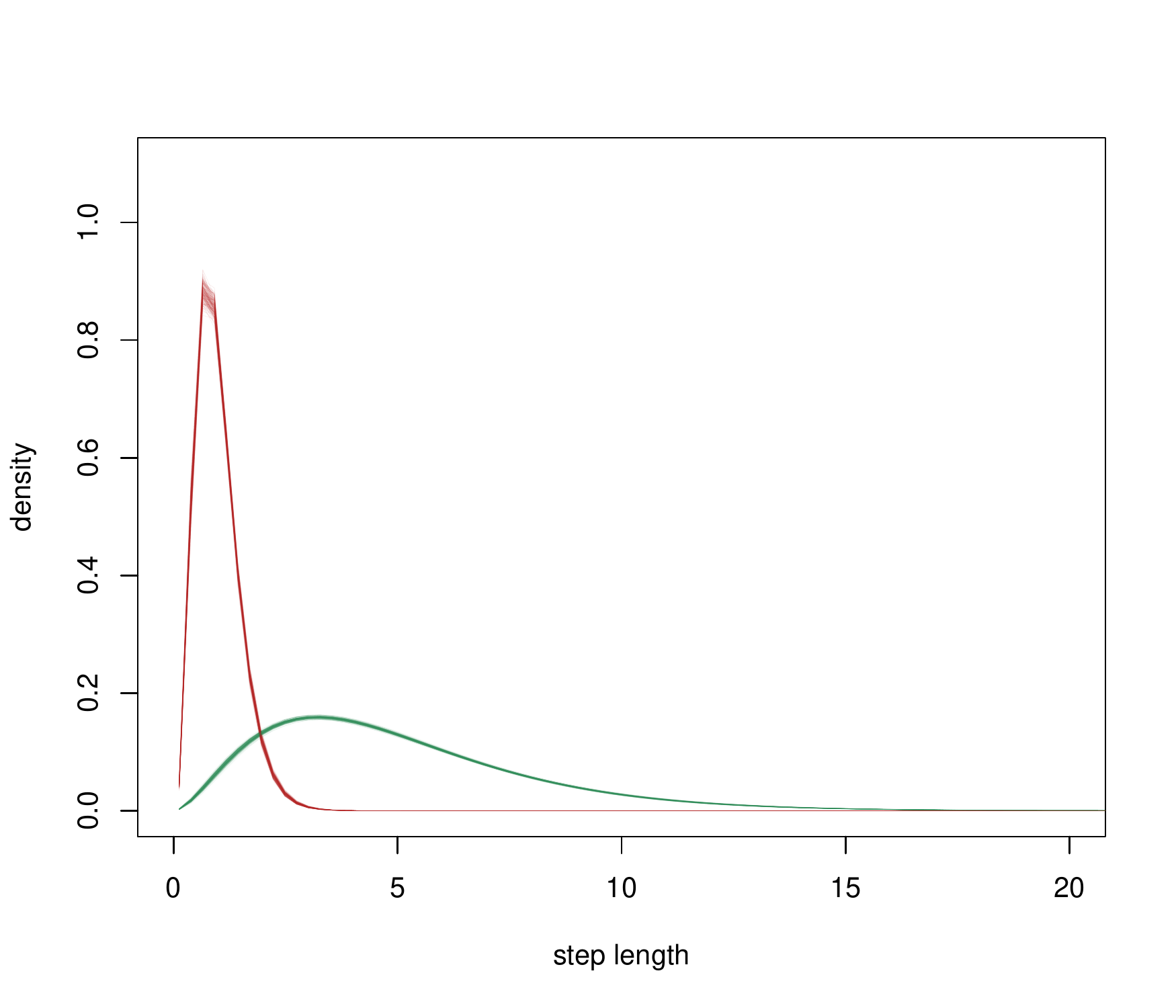} \includegraphics[width=0.49\linewidth]{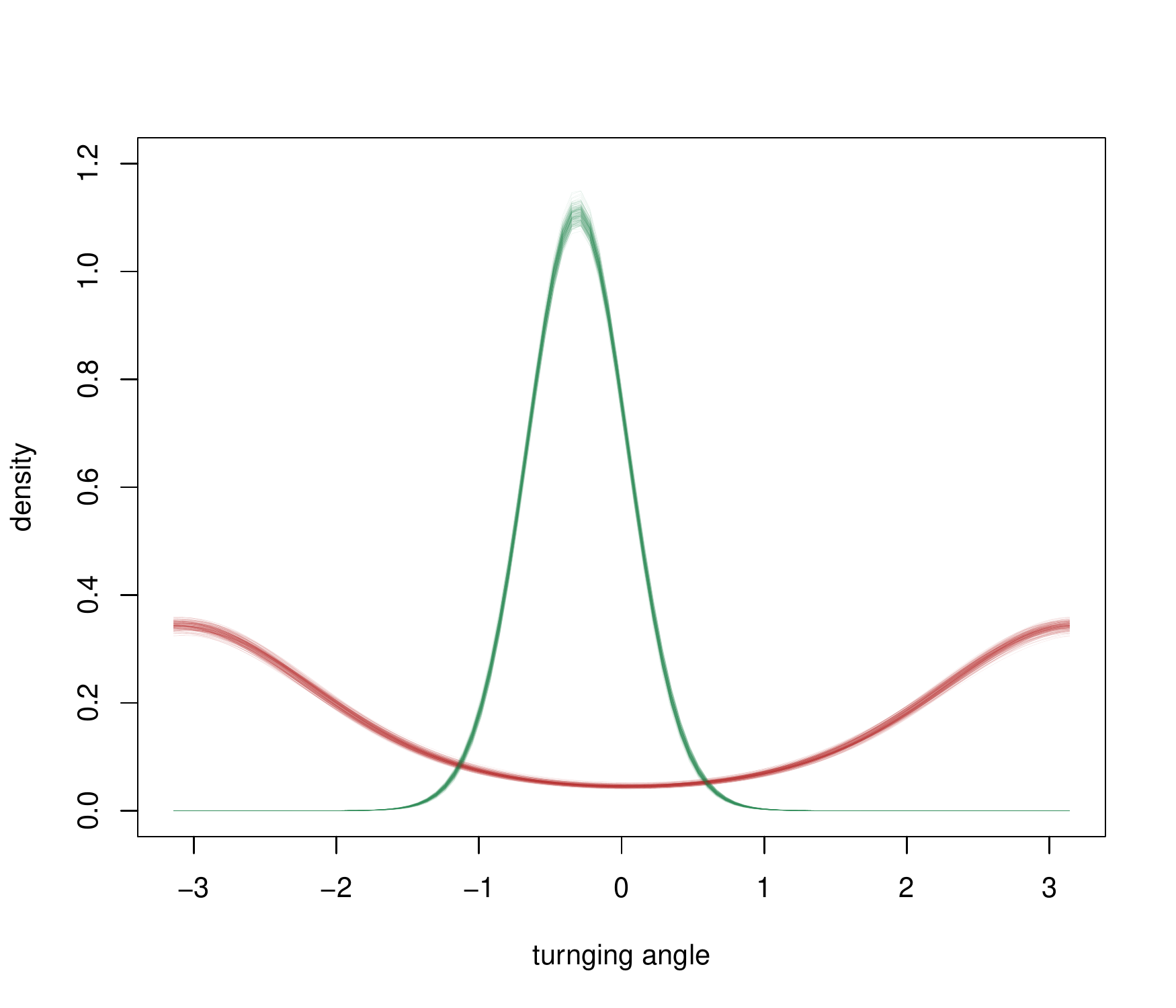} 

}

\caption{Histograms of the observed step lengths (left) and turning angles (right), with the estimated state-dependent densities weighted by the stationary distributions.}\label{fig:unnamed-chunk-21}
\end{figure}

We can also plot the transition probabilities as functions of the
covariates. For example, we use the following code to visualise the
effect of the slope on the transition probabilities when temperature is
equal to 10.

\begin{Shaded}
\begin{Highlighting}[]
\CommentTok{# extract parameters of the t.p.m}
\NormalTok{samp <-}\StringTok{ }\KeywordTok{as.matrix}\NormalTok{(fit)}
\NormalTok{beta <-}\StringTok{ }\NormalTok{samp[, }\KeywordTok{grep}\NormalTok{(}\StringTok{"beta"}\NormalTok{, }\KeywordTok{colnames}\NormalTok{(samp))]}

\CommentTok{# build a design matrix}
\NormalTok{gridslope <-}\StringTok{ }\KeywordTok{seq}\NormalTok{(}\KeywordTok{min}\NormalTok{(data}\OperatorTok{$}\NormalTok{slope), }\KeywordTok{max}\NormalTok{(data}\OperatorTok{$}\NormalTok{slope), }\DataTypeTok{length =} \DecValTok{100}\NormalTok{)}
\NormalTok{gridslopesc <-}\StringTok{ }\KeywordTok{seq}\NormalTok{(}\KeywordTok{min}\NormalTok{(}\KeywordTok{scale}\NormalTok{(data}\OperatorTok{$}\NormalTok{slope)), }\KeywordTok{max}\NormalTok{(}\KeywordTok{scale}\NormalTok{(data}\OperatorTok{$}\NormalTok{slope)), }\DataTypeTok{length =} \DecValTok{100}\NormalTok{)}
\NormalTok{fixedtemp <-}\StringTok{ }\DecValTok{10}
\NormalTok{DM <-}\StringTok{ }\KeywordTok{cbind}\NormalTok{(}\DecValTok{1}\NormalTok{, fixedtemp, gridslopesc, gridslopesc}\OperatorTok{^}\DecValTok{2}\NormalTok{)}

\CommentTok{# indices of posterior draws to plot (thinned for visualisation purposes)}
\NormalTok{ind <-}\StringTok{ }\KeywordTok{seq}\NormalTok{(}\DecValTok{1}\NormalTok{, }\KeywordTok{nrow}\NormalTok{(samp), }\DataTypeTok{by =} \DecValTok{5}\NormalTok{)}

\CommentTok{# plot the transition probabilities}
\KeywordTok{par}\NormalTok{(}\DataTypeTok{mfrow =} \KeywordTok{c}\NormalTok{(}\DecValTok{2}\NormalTok{, }\DecValTok{2}\NormalTok{))}
\ControlFlowTok{for}\NormalTok{ (i }\ControlFlowTok{in} \DecValTok{1}\OperatorTok{:}\DecValTok{2}\NormalTok{) \{}
    \ControlFlowTok{for}\NormalTok{ (j }\ControlFlowTok{in} \DecValTok{1}\OperatorTok{:}\DecValTok{2}\NormalTok{) \{}
\NormalTok{        tpm <-}\StringTok{ }\NormalTok{moveHMM}\OperatorTok{:::}\KeywordTok{trMatrix_rcpp}\NormalTok{(}\DataTypeTok{nbStates =} \DecValTok{2}\NormalTok{, }\DataTypeTok{beta =} \KeywordTok{t}\NormalTok{(}\KeywordTok{matrix}\NormalTok{(beta[ind[}\DecValTok{1}\NormalTok{], }
\NormalTok{            ], }\DataTypeTok{ncol =} \KeywordTok{ncol}\NormalTok{(DM))), }\DataTypeTok{covs =}\NormalTok{ DM)}
        \KeywordTok{plot}\NormalTok{(gridslope, tpm[i, j, ], }\DataTypeTok{type =} \StringTok{"l"}\NormalTok{, }\DataTypeTok{ylim =} \KeywordTok{c}\NormalTok{(}\DecValTok{0}\NormalTok{, }\DecValTok{1}\NormalTok{), }\DataTypeTok{col =} \KeywordTok{rgb}\NormalTok{(}\DecValTok{1}\NormalTok{, }
            \DecValTok{0}\NormalTok{, }\DecValTok{0}\NormalTok{, }\FloatTok{0.3}\NormalTok{), }\DataTypeTok{lwd =} \FloatTok{0.5}\NormalTok{, }\DataTypeTok{xlab =} \StringTok{"slope"}\NormalTok{, }\DataTypeTok{ylab =} \KeywordTok{paste0}\NormalTok{(}\StringTok{"Pr("}\NormalTok{, i, }\StringTok{" -> "}\NormalTok{, }
\NormalTok{            j, }\StringTok{")"}\NormalTok{))}
        \ControlFlowTok{for}\NormalTok{ (k }\ControlFlowTok{in}\NormalTok{ ind[}\OperatorTok{-}\DecValTok{1}\NormalTok{]) \{}
\NormalTok{            tpm <-}\StringTok{ }\NormalTok{moveHMM}\OperatorTok{:::}\KeywordTok{trMatrix_rcpp}\NormalTok{(}\DataTypeTok{nbStates =} \DecValTok{2}\NormalTok{, }\DataTypeTok{beta =} \KeywordTok{t}\NormalTok{(}\KeywordTok{matrix}\NormalTok{(beta[k, }
\NormalTok{                ], }\DataTypeTok{ncol =} \KeywordTok{ncol}\NormalTok{(DM))), }\DataTypeTok{covs =}\NormalTok{ DM)}
            \KeywordTok{points}\NormalTok{(gridslope, tpm[i, j, ], }\DataTypeTok{type =} \StringTok{"l"}\NormalTok{, }\DataTypeTok{col =} \KeywordTok{rgb}\NormalTok{(}\DecValTok{0}\NormalTok{, }\DecValTok{0}\NormalTok{, }\DecValTok{0}\NormalTok{, }\FloatTok{0.2}\NormalTok{), }
                \DataTypeTok{lwd =} \FloatTok{0.5}\NormalTok{)}
\NormalTok{        \}}
\NormalTok{    \}}
\NormalTok{\}}
\end{Highlighting}
\end{Shaded}

\begin{figure}

{\centering \includegraphics[width=0.8\linewidth]{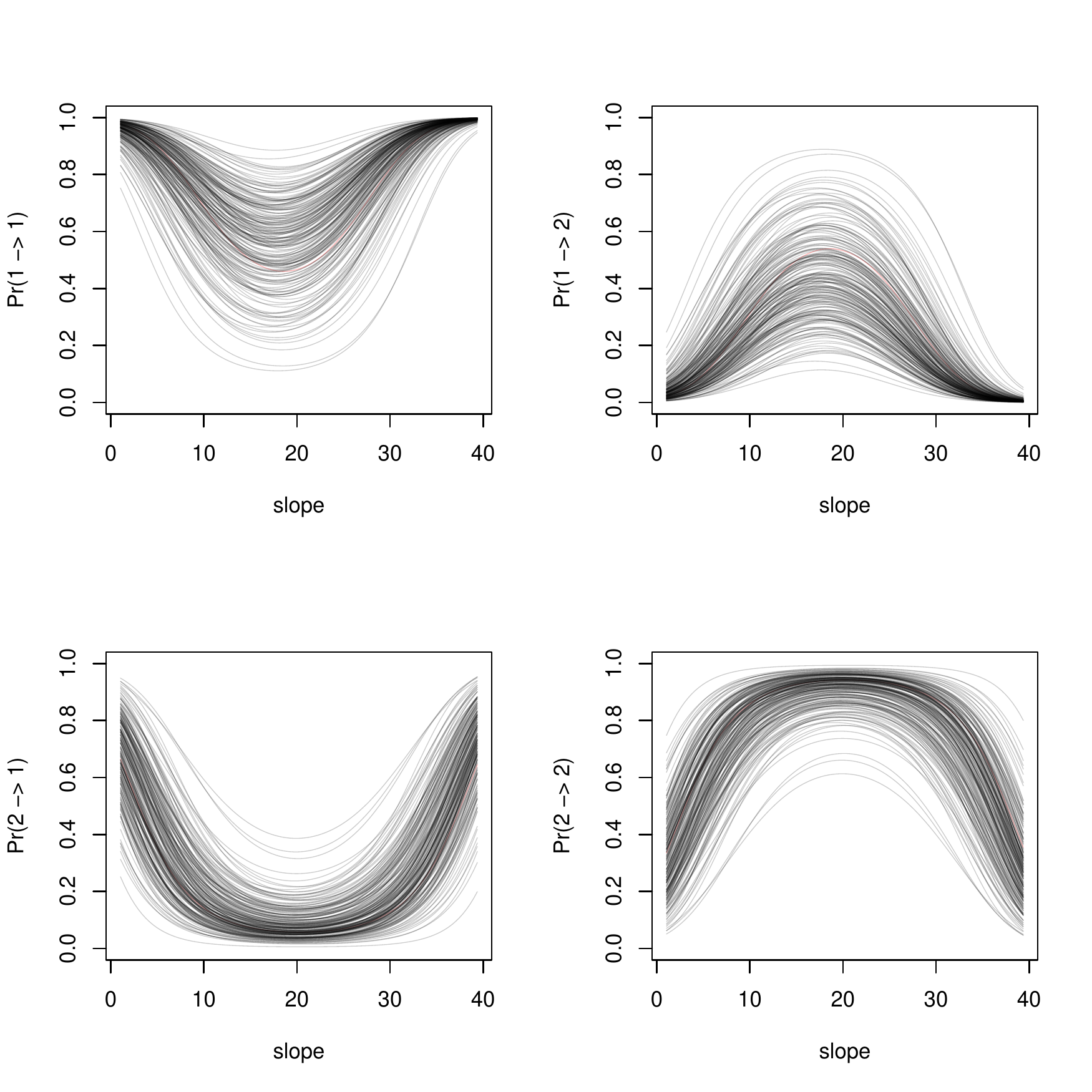} 

}

\caption{Posterior transition probabilities as functions of the slope covariate.}\label{fig:unnamed-chunk-22}
\end{figure}

\begin{Shaded}
\begin{Highlighting}[]
\KeywordTok{par}\NormalTok{(}\DataTypeTok{mfrow =} \KeywordTok{c}\NormalTok{(}\DecValTok{1}\NormalTok{, }\DecValTok{1}\NormalTok{))}
\end{Highlighting}
\end{Shaded}

We perform the same graphical posterior predictive checks from before.
First, we simulate data using draws from the posterior distribution of
the parameters:

\begin{Shaded}
\begin{Highlighting}[]
\NormalTok{## generate new data sets}

\NormalTok{n.sims <-}\StringTok{ }\KeywordTok{dim}\NormalTok{(kappa)[}\DecValTok{1}\NormalTok{]}
\NormalTok{n <-}\StringTok{ }\KeywordTok{length}\NormalTok{(data}\OperatorTok{$}\NormalTok{step)}

\CommentTok{# state sequences}
\NormalTok{ppstates <-}\StringTok{ }\KeywordTok{matrix}\NormalTok{(}\OtherTok{NA}\NormalTok{, }\DataTypeTok{nrow =}\NormalTok{ n.sims, }\DataTypeTok{ncol =}\NormalTok{ n)}
\CommentTok{# observations}
\NormalTok{ppsteps <-}\StringTok{ }\KeywordTok{matrix}\NormalTok{(}\OtherTok{NA}\NormalTok{, }\DataTypeTok{nrow =}\NormalTok{ n.sims, }\DataTypeTok{ncol =}\NormalTok{ n)}
\NormalTok{ppangs <-}\StringTok{ }\KeywordTok{matrix}\NormalTok{(}\OtherTok{NA}\NormalTok{, }\DataTypeTok{nrow =}\NormalTok{ n.sims, }\DataTypeTok{ncol =}\NormalTok{ n)}


\NormalTok{DM <-}\StringTok{ }\KeywordTok{cbind}\NormalTok{(}\DecValTok{1}\NormalTok{, }\KeywordTok{scale}\NormalTok{(data}\OperatorTok{$}\NormalTok{temp), }\KeywordTok{scale}\NormalTok{(data}\OperatorTok{$}\NormalTok{slope), }\KeywordTok{scale}\NormalTok{(data}\OperatorTok{$}\NormalTok{slope)}\OperatorTok{^}\DecValTok{2}\NormalTok{)}
\ControlFlowTok{for}\NormalTok{ (j }\ControlFlowTok{in} \DecValTok{1}\OperatorTok{:}\NormalTok{n.sims) \{}
\NormalTok{    tpm <-}\StringTok{ }\NormalTok{moveHMM}\OperatorTok{:::}\KeywordTok{trMatrix_rcpp}\NormalTok{(}\DataTypeTok{nbStates =} \DecValTok{2}\NormalTok{, }\DataTypeTok{beta =} \KeywordTok{t}\NormalTok{(}\KeywordTok{matrix}\NormalTok{(beta[j, ], }
        \DataTypeTok{ncol =} \KeywordTok{ncol}\NormalTok{(DM))), }\DataTypeTok{covs =}\NormalTok{ DM)}
\NormalTok{    initdist <-}\StringTok{ }\KeywordTok{rep}\NormalTok{(}\DecValTok{1}\OperatorTok{/}\NormalTok{N, N)}
    
\NormalTok{    ppstates[j, }\DecValTok{1}\NormalTok{] <-}\StringTok{ }\KeywordTok{sample}\NormalTok{(}\DecValTok{1}\OperatorTok{:}\NormalTok{N, }\DataTypeTok{size =} \DecValTok{1}\NormalTok{, }\DataTypeTok{prob =}\NormalTok{ initdist)}
\NormalTok{    ppsteps[j, }\DecValTok{1}\NormalTok{] <-}\StringTok{ }\KeywordTok{rgamma}\NormalTok{(}\DecValTok{1}\NormalTok{, }\DataTypeTok{shape =}\NormalTok{ shape[j, ppstates[j, }\DecValTok{1}\NormalTok{]], }\DataTypeTok{rate =}\NormalTok{ rate[j, }
\NormalTok{        ppstates[j, }\DecValTok{1}\NormalTok{]])}
\NormalTok{    ppangs[j, }\DecValTok{1}\NormalTok{] <-}\StringTok{ }\KeywordTok{rvm}\NormalTok{(}\DecValTok{1}\NormalTok{, }\DataTypeTok{mean =}\NormalTok{ loc[j, ppstates[j, }\DecValTok{1}\NormalTok{]], }\DataTypeTok{k =}\NormalTok{ kappa[j, ppstates[j, }
        \DecValTok{1}\NormalTok{]])}
    
    \ControlFlowTok{for}\NormalTok{ (i }\ControlFlowTok{in} \DecValTok{2}\OperatorTok{:}\NormalTok{n) \{}
\NormalTok{        ppstates[j, i] <-}\StringTok{ }\KeywordTok{sample}\NormalTok{(}\DecValTok{1}\OperatorTok{:}\NormalTok{N, }\DataTypeTok{size =} \DecValTok{1}\NormalTok{, }\DataTypeTok{prob =}\NormalTok{ tpm[ppstates[j, i }\OperatorTok{-}\StringTok{ }\DecValTok{1}\NormalTok{], }
\NormalTok{            , i])}
\NormalTok{        ppsteps[j, i] <-}\StringTok{ }\KeywordTok{rgamma}\NormalTok{(}\DecValTok{1}\NormalTok{, }\DataTypeTok{shape =}\NormalTok{ shape[j, ppstates[j, i]], }\DataTypeTok{rate =}\NormalTok{ rate[j, }
\NormalTok{            ppstates[j, i]])}
\NormalTok{        ppangs[j, i] <-}\StringTok{ }\KeywordTok{rvm}\NormalTok{(}\DecValTok{1}\NormalTok{, }\DataTypeTok{mean =}\NormalTok{ loc[j, ppstates[j, i]], }\DataTypeTok{k =}\NormalTok{ kappa[j, ppstates[j, }
\NormalTok{            i]])}
\NormalTok{    \}}
\NormalTok{\}}

\ControlFlowTok{for}\NormalTok{ (j }\ControlFlowTok{in} \DecValTok{1}\OperatorTok{:}\NormalTok{n.sims) ppangs[j, ] <-}\StringTok{ }\KeywordTok{as.numeric}\NormalTok{(}\KeywordTok{minusPiPlusPi}\NormalTok{(ppangs[j, ]))}
\end{Highlighting}
\end{Shaded}

We check that the densities of the replicated data sets are similar to
the observed data set, for both step lengths and turning angles.

\begin{Shaded}
\begin{Highlighting}[]
\KeywordTok{ppc_dens_overlay}\NormalTok{(data}\OperatorTok{$}\NormalTok{step[}\KeywordTok{which}\NormalTok{(}\OperatorTok{!}\KeywordTok{is.na}\NormalTok{(data}\OperatorTok{$}\NormalTok{step))], }
\NormalTok{                 ppsteps[}\DecValTok{1}\OperatorTok{:}\DecValTok{100}\NormalTok{,}\KeywordTok{which}\NormalTok{(}\OperatorTok{!}\KeywordTok{is.na}\NormalTok{(data}\OperatorTok{$}\NormalTok{step))])}
\end{Highlighting}
\end{Shaded}

\begin{center}\includegraphics[width=0.49\linewidth]{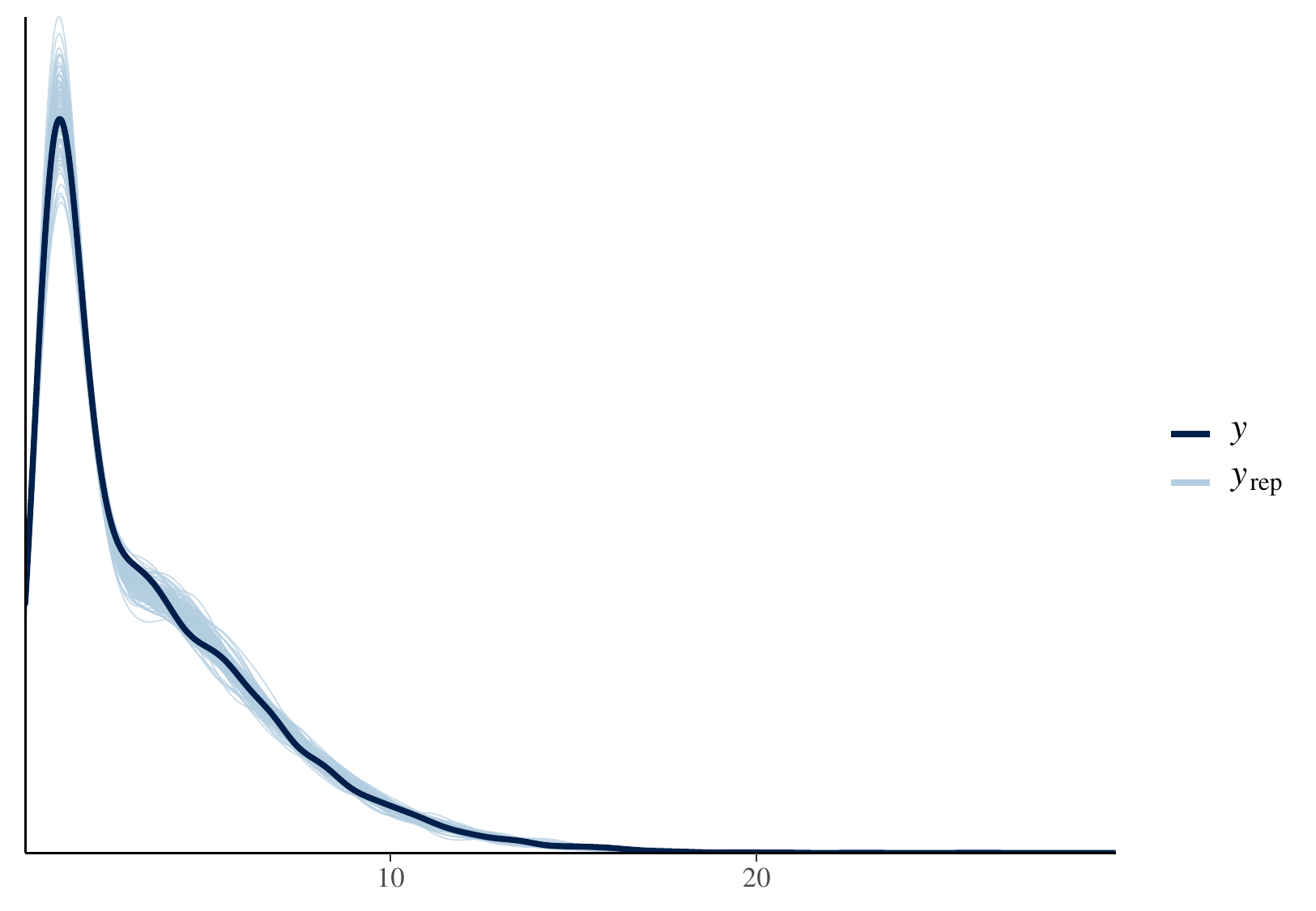} \end{center}

\begin{Shaded}
\begin{Highlighting}[]
\KeywordTok{ppc_dens_overlay}\NormalTok{(data}\OperatorTok{$}\NormalTok{angle[}\KeywordTok{which}\NormalTok{(}\OperatorTok{!}\KeywordTok{is.na}\NormalTok{(data}\OperatorTok{$}\NormalTok{angle))], }
\NormalTok{                 ppangs[}\DecValTok{1}\OperatorTok{:}\DecValTok{100}\NormalTok{,}\KeywordTok{which}\NormalTok{(}\OperatorTok{!}\KeywordTok{is.na}\NormalTok{(data}\OperatorTok{$}\NormalTok{angle))])}
\end{Highlighting}
\end{Shaded}

\begin{center}\includegraphics[width=0.49\linewidth]{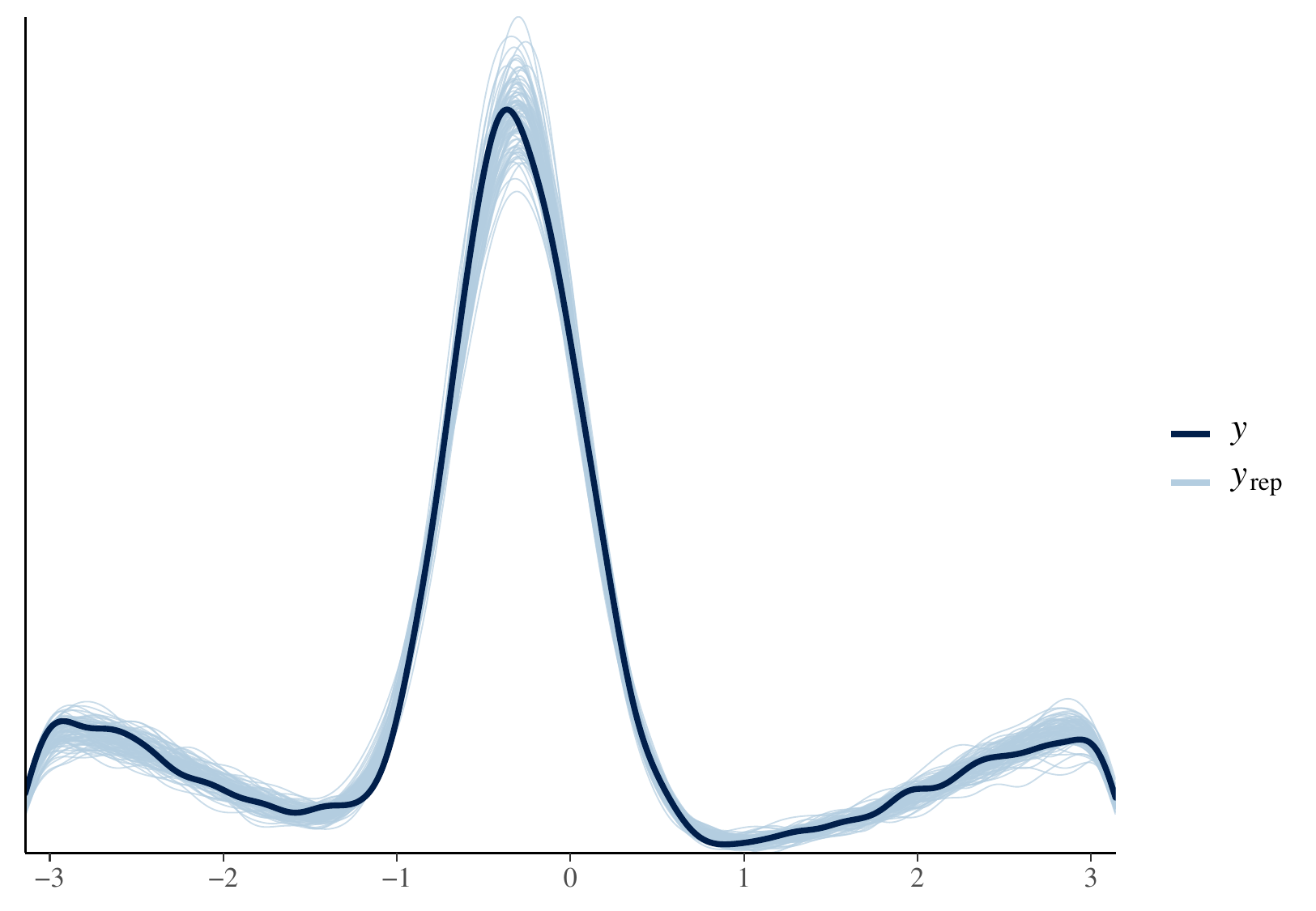} \end{center}

We compare the observed autocorrelation with the autocorrelation of the
simulated data sets.

\begin{Shaded}
\begin{Highlighting}[]
\NormalTok{nlags <-}\StringTok{ }\DecValTok{61}
\CommentTok{# observed acf}
\NormalTok{oac =}\StringTok{ }\KeywordTok{acf}\NormalTok{(data}\OperatorTok{$}\NormalTok{step[}\DecValTok{2}\OperatorTok{:}\NormalTok{(n }\OperatorTok{-}\StringTok{ }\DecValTok{1}\NormalTok{)], }\DataTypeTok{lag.max =}\NormalTok{ (nlags }\OperatorTok{-}\StringTok{ }\DecValTok{1}\NormalTok{), }\DataTypeTok{plot =} \OtherTok{FALSE}\NormalTok{, }\DataTypeTok{na.action =}\NormalTok{ na.pass)}\OperatorTok{$}\NormalTok{acf}
\NormalTok{ppac =}\StringTok{ }\KeywordTok{matrix}\NormalTok{(}\OtherTok{NA}\NormalTok{, n.sims, nlags)}
\ControlFlowTok{for}\NormalTok{ (i }\ControlFlowTok{in} \DecValTok{1}\OperatorTok{:}\NormalTok{n.sims) \{}
\NormalTok{    ppac[i, ] =}\StringTok{ }\KeywordTok{acf}\NormalTok{(ppsteps[i, ], }\DataTypeTok{lag.max =}\NormalTok{ (nlags }\OperatorTok{-}\StringTok{ }\DecValTok{1}\NormalTok{), }\DataTypeTok{plot =} \OtherTok{FALSE}\NormalTok{)}\OperatorTok{$}\NormalTok{acf}
\NormalTok{\}}

\NormalTok{hpd.acf <-}\StringTok{ }\KeywordTok{HPDinterval}\NormalTok{(}\KeywordTok{as.mcmc}\NormalTok{(ppac), }\DataTypeTok{prob =} \FloatTok{0.95}\NormalTok{)}
\NormalTok{dat <-}\StringTok{ }\KeywordTok{data.frame}\NormalTok{(}\DataTypeTok{y =} \DecValTok{1}\OperatorTok{:}\DecValTok{61}\NormalTok{, }\DataTypeTok{acf =} \KeywordTok{as.numeric}\NormalTok{(oac), }\DataTypeTok{lb =}\NormalTok{ hpd.acf[, }\DecValTok{1}\NormalTok{], }\DataTypeTok{ub =}\NormalTok{ hpd.acf[, }
    \DecValTok{2}\NormalTok{])}

\KeywordTok{ggplot}\NormalTok{(dat, }\KeywordTok{aes}\NormalTok{(y, acf)) }\OperatorTok{+}\StringTok{ }\KeywordTok{geom_ribbon}\NormalTok{(}\KeywordTok{aes}\NormalTok{(}\DataTypeTok{x =}\NormalTok{ y, }\DataTypeTok{ymin =}\NormalTok{ lb, }\DataTypeTok{ymax =}\NormalTok{ ub), }\DataTypeTok{fill =} \StringTok{"grey70"}\NormalTok{, }
    \DataTypeTok{alpha =} \FloatTok{0.5}\NormalTok{) }\OperatorTok{+}\StringTok{ }\KeywordTok{geom_point}\NormalTok{(}\DataTypeTok{col =} \StringTok{"purple"}\NormalTok{, }\DataTypeTok{size =} \DecValTok{1}\NormalTok{) }\OperatorTok{+}\StringTok{ }\KeywordTok{geom_line}\NormalTok{() }\OperatorTok{+}\StringTok{ }\KeywordTok{coord_cartesian}\NormalTok{(}\DataTypeTok{xlim =} \KeywordTok{c}\NormalTok{(}\DecValTok{2}\NormalTok{, }
    \DecValTok{60}\NormalTok{), }\DataTypeTok{ylim =} \KeywordTok{c}\NormalTok{(}\OperatorTok{-}\FloatTok{0.1}\NormalTok{, }\FloatTok{0.5}\NormalTok{)) }\OperatorTok{+}\StringTok{ }\KeywordTok{xlab}\NormalTok{(}\StringTok{"Lag"}\NormalTok{) }\OperatorTok{+}\StringTok{ }\KeywordTok{ylab}\NormalTok{(}\StringTok{"ACF"}\NormalTok{) }\OperatorTok{+}\StringTok{ }\KeywordTok{ggtitle}\NormalTok{(}\StringTok{"Observed Autocorrelation Function}
\StringTok{   with 90% CI for ACF of Predicted Quantities"}\NormalTok{)}
\end{Highlighting}
\end{Shaded}

\begin{center}\includegraphics[width=0.7\linewidth]{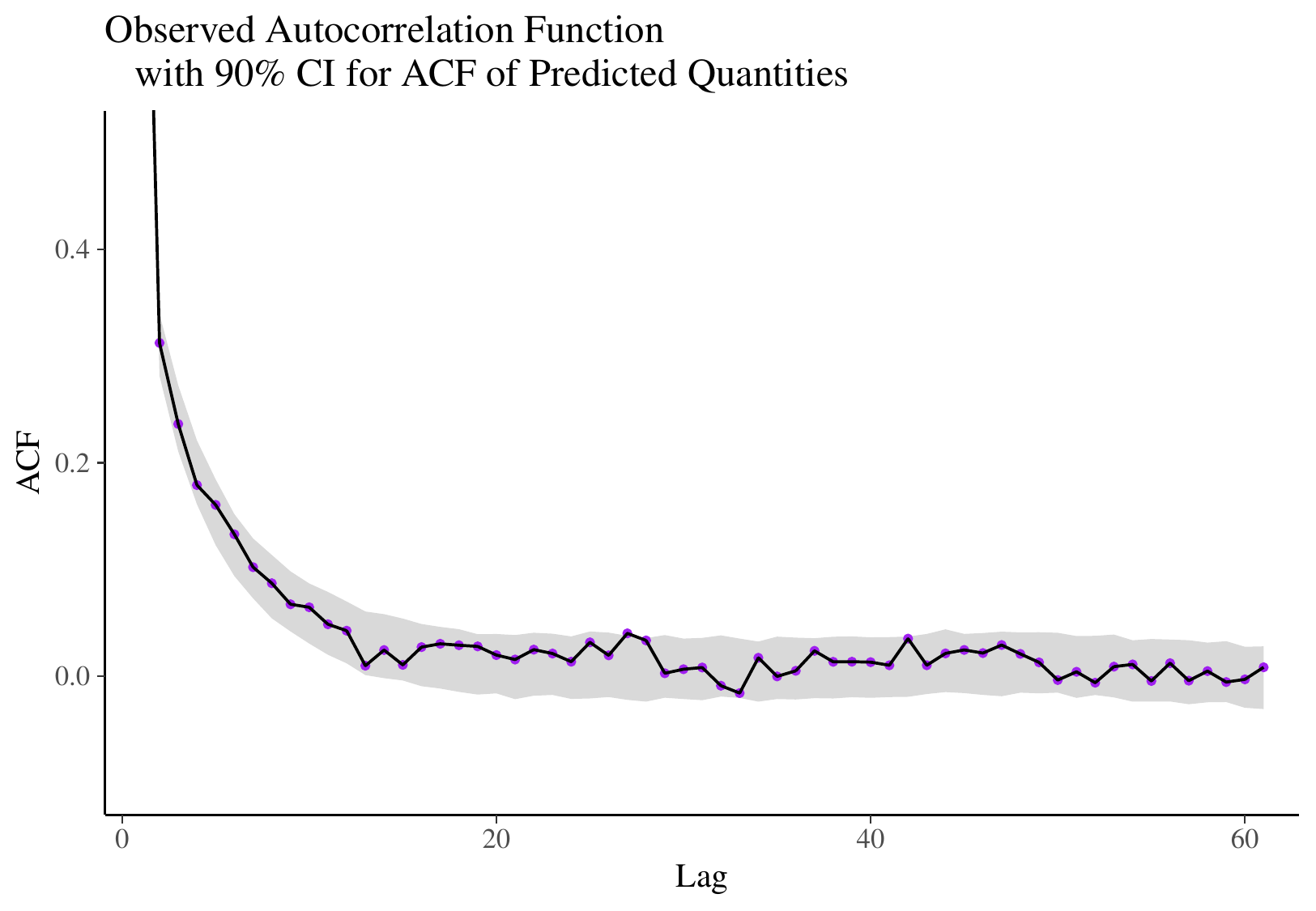} \end{center}

We also compute the forecast (pseudo-)residuals for the step lengths. We
make an adjustment to the previous code because the t.p.m.~is no longer
stationary.

\begin{Shaded}
\begin{Highlighting}[]
\NormalTok{##### R code for Forecast (Pseudo-)Residuals}

\NormalTok{## Calculating forward variables}
\NormalTok{HMM.lalpha <-}\StringTok{ }\ControlFlowTok{function}\NormalTok{(allprobs, gamma, delta, n, N) \{}
\NormalTok{    lalpha <-}\StringTok{ }\KeywordTok{matrix}\NormalTok{(}\OtherTok{NA}\NormalTok{, N, n)}
    
\NormalTok{    lscale <-}\StringTok{ }\DecValTok{0}
\NormalTok{    foo <-}\StringTok{ }\NormalTok{delta }\OperatorTok{*}\StringTok{ }\NormalTok{allprobs[}\DecValTok{1}\NormalTok{, ]}
\NormalTok{    lscale <-}\StringTok{ }\DecValTok{0}
\NormalTok{    lalpha[, }\DecValTok{1}\NormalTok{] <-}\StringTok{ }\KeywordTok{log}\NormalTok{(foo) }\OperatorTok{+}\StringTok{ }\NormalTok{lscale}
\NormalTok{    sumfoo <-}\StringTok{ }\KeywordTok{sum}\NormalTok{(foo)}
    \ControlFlowTok{for}\NormalTok{ (i }\ControlFlowTok{in} \DecValTok{2}\OperatorTok{:}\NormalTok{n) \{}
\NormalTok{        foo <-}\StringTok{ }\NormalTok{foo }\OperatorTok{%*%}\StringTok{ }\NormalTok{gamma[, , i] }\OperatorTok{*}\StringTok{ }\NormalTok{allprobs[i, ]}
        \CommentTok{# scaling}
\NormalTok{        sumfoo <-}\StringTok{ }\KeywordTok{sum}\NormalTok{(foo)}
\NormalTok{        lscale <-}\StringTok{ }\NormalTok{lscale }\OperatorTok{+}\StringTok{ }\KeywordTok{log}\NormalTok{(sumfoo)}
\NormalTok{        foo <-}\StringTok{ }\NormalTok{foo}\OperatorTok{/}\NormalTok{sumfoo}
\NormalTok{        lalpha[, i] <-}\StringTok{ }\KeywordTok{log}\NormalTok{(foo) }\OperatorTok{+}\StringTok{ }\NormalTok{lscale}
\NormalTok{    \}}
\NormalTok{    lalpha}
\NormalTok{\}}


\NormalTok{## Calculating forecast (pseudo-)residuals}
\NormalTok{HMM.psres <-}\StringTok{ }\ControlFlowTok{function}\NormalTok{(x, allprobs, gamma, n, N, shape, rate) \{}
    
    
\NormalTok{    delta <-}\StringTok{ }\KeywordTok{rep}\NormalTok{(}\DecValTok{1}\OperatorTok{/}\NormalTok{N, N)}
    
\NormalTok{    la <-}\StringTok{ }\KeywordTok{HMM.lalpha}\NormalTok{(allprobs, gamma, delta, n, N)}
    
\NormalTok{    pstepmat <-}\StringTok{ }\KeywordTok{matrix}\NormalTok{(}\OtherTok{NA}\NormalTok{, n, N)}
\NormalTok{    fres <-}\StringTok{ }\KeywordTok{rep}\NormalTok{(}\OtherTok{NA}\NormalTok{, n)}
\NormalTok{    ind.step <-}\StringTok{ }\KeywordTok{which}\NormalTok{(}\OperatorTok{!}\KeywordTok{is.na}\NormalTok{(x))}
    
    \ControlFlowTok{for}\NormalTok{ (j }\ControlFlowTok{in} \DecValTok{1}\OperatorTok{:}\KeywordTok{length}\NormalTok{(ind.step)) \{}
\NormalTok{        pstepmat[ind.step[j], }\DecValTok{1}\NormalTok{] <-}\StringTok{ }\KeywordTok{pgamma}\NormalTok{(x[ind.step[j]], }\DataTypeTok{shape =}\NormalTok{ shape[}\DecValTok{1}\NormalTok{], }
            \DataTypeTok{rate =}\NormalTok{ rate[}\DecValTok{1}\NormalTok{])}
\NormalTok{        pstepmat[ind.step[j], }\DecValTok{2}\NormalTok{] <-}\StringTok{ }\KeywordTok{pgamma}\NormalTok{(x[ind.step[j]], }\DataTypeTok{shape =}\NormalTok{ shape[}\DecValTok{2}\NormalTok{], }
            \DataTypeTok{rate =}\NormalTok{ rate[}\DecValTok{2}\NormalTok{])}
\NormalTok{    \}}
    
    
    \ControlFlowTok{if}\NormalTok{ (}\OperatorTok{!}\KeywordTok{is.na}\NormalTok{(x[}\DecValTok{1}\NormalTok{])) }
\NormalTok{        fres[}\DecValTok{1}\NormalTok{] <-}\StringTok{ }\KeywordTok{qnorm}\NormalTok{(}\KeywordTok{rbind}\NormalTok{(}\KeywordTok{c}\NormalTok{(}\DecValTok{1}\NormalTok{, }\DecValTok{0}\NormalTok{)) }\OperatorTok{%*%}\StringTok{ }\NormalTok{pstepmat[}\DecValTok{1}\NormalTok{, ])}
    \ControlFlowTok{for}\NormalTok{ (i }\ControlFlowTok{in} \DecValTok{2}\OperatorTok{:}\NormalTok{n) \{}
        
\NormalTok{        c <-}\StringTok{ }\KeywordTok{max}\NormalTok{(la[, i }\OperatorTok{-}\StringTok{ }\DecValTok{1}\NormalTok{])}
\NormalTok{        a <-}\StringTok{ }\KeywordTok{exp}\NormalTok{(la[, i }\OperatorTok{-}\StringTok{ }\DecValTok{1}\NormalTok{] }\OperatorTok{-}\StringTok{ }\NormalTok{c)}
        \ControlFlowTok{if}\NormalTok{ (}\OperatorTok{!}\KeywordTok{is.na}\NormalTok{(x[i])) }
\NormalTok{            fres[i] <-}\StringTok{ }\KeywordTok{qnorm}\NormalTok{(}\KeywordTok{t}\NormalTok{(a) }\OperatorTok{%*%}\StringTok{ }\NormalTok{(gamma[, , i]}\OperatorTok{/}\KeywordTok{sum}\NormalTok{(a)) }\OperatorTok{%*%}\StringTok{ }\NormalTok{pstepmat[i, }
\NormalTok{                ])}
\NormalTok{    \}}
    \KeywordTok{return}\NormalTok{(}\KeywordTok{list}\NormalTok{(}\DataTypeTok{fres =}\NormalTok{ fres))}
\NormalTok{\}}
\end{Highlighting}
\end{Shaded}

\begin{Shaded}
\begin{Highlighting}[]
\NormalTok{shape.est <-}\StringTok{ }\KeywordTok{colMeans}\NormalTok{(shape)}
\NormalTok{rate.est <-}\StringTok{ }\KeywordTok{colMeans}\NormalTok{(rate)}
\NormalTok{beta.est <-}\StringTok{ }\KeywordTok{colMeans}\NormalTok{(beta)}

\NormalTok{allprobs <-}\StringTok{ }\KeywordTok{matrix}\NormalTok{(}\DecValTok{1}\NormalTok{, }\DataTypeTok{nrow =}\NormalTok{ n, }\DataTypeTok{ncol =}\NormalTok{ N)}
\ControlFlowTok{for}\NormalTok{ (j }\ControlFlowTok{in} \DecValTok{1}\OperatorTok{:}\NormalTok{N) allprobs[}\KeywordTok{which}\NormalTok{(}\OperatorTok{!}\KeywordTok{is.na}\NormalTok{(data}\OperatorTok{$}\NormalTok{step)), j] <-}\StringTok{ }\KeywordTok{dgamma}\NormalTok{(data[}\KeywordTok{which}\NormalTok{(}\OperatorTok{!}\KeywordTok{is.na}\NormalTok{(data}\OperatorTok{$}\NormalTok{step)), }
    \StringTok{"step"}\NormalTok{], }\DataTypeTok{shape =}\NormalTok{ shape.est, }\DataTypeTok{rate =}\NormalTok{ rate.est)}

\NormalTok{gamma <-}\StringTok{ }\NormalTok{moveHMM}\OperatorTok{:::}\KeywordTok{trMatrix_rcpp}\NormalTok{(}\DataTypeTok{nbStates =} \DecValTok{2}\NormalTok{, }\DataTypeTok{beta =} \KeywordTok{t}\NormalTok{(}\KeywordTok{matrix}\NormalTok{(beta.est, }\DataTypeTok{ncol =} \KeywordTok{ncol}\NormalTok{(DM))), }
    \DataTypeTok{covs =}\NormalTok{ DM)}

\NormalTok{fres <-}\StringTok{ }\KeywordTok{HMM.psres}\NormalTok{(}\DataTypeTok{x =}\NormalTok{ data}\OperatorTok{$}\NormalTok{step, }\DataTypeTok{allprobs =}\NormalTok{ allprobs, }\DataTypeTok{gamma =}\NormalTok{ gamma, }\DataTypeTok{n =}\NormalTok{ n, }
    \DataTypeTok{N =}\NormalTok{ N, }\DataTypeTok{shape =}\NormalTok{ shape.est, }\DataTypeTok{rate =}\NormalTok{ rate.est)}
\end{Highlighting}
\end{Shaded}

Plotting the residuals in a Q-Q plot:

\begin{Shaded}
\begin{Highlighting}[]
\KeywordTok{ggplot}\NormalTok{(}\DataTypeTok{data=}\KeywordTok{data.frame}\NormalTok{(}\DataTypeTok{x=}\NormalTok{fres}\OperatorTok{$}\NormalTok{fres), }\KeywordTok{aes}\NormalTok{(}\DataTypeTok{sample =}\NormalTok{ x)) }\OperatorTok{+}\StringTok{ }\KeywordTok{stat_qq}\NormalTok{() }\OperatorTok{+}\StringTok{ }
\StringTok{  }\KeywordTok{stat_qq_line}\NormalTok{(}\DataTypeTok{color=}\StringTok{"purple"}\NormalTok{, }\DataTypeTok{size=}\DecValTok{1}\NormalTok{) }\OperatorTok{+}
\StringTok{  }\KeywordTok{ggtitle}\NormalTok{(}\StringTok{"Q-Q Plot"}\NormalTok{) }\OperatorTok{+}\StringTok{ }\KeywordTok{theme_classic}\NormalTok{()}
\end{Highlighting}
\end{Shaded}

\begin{center}\includegraphics[width=0.7\linewidth]{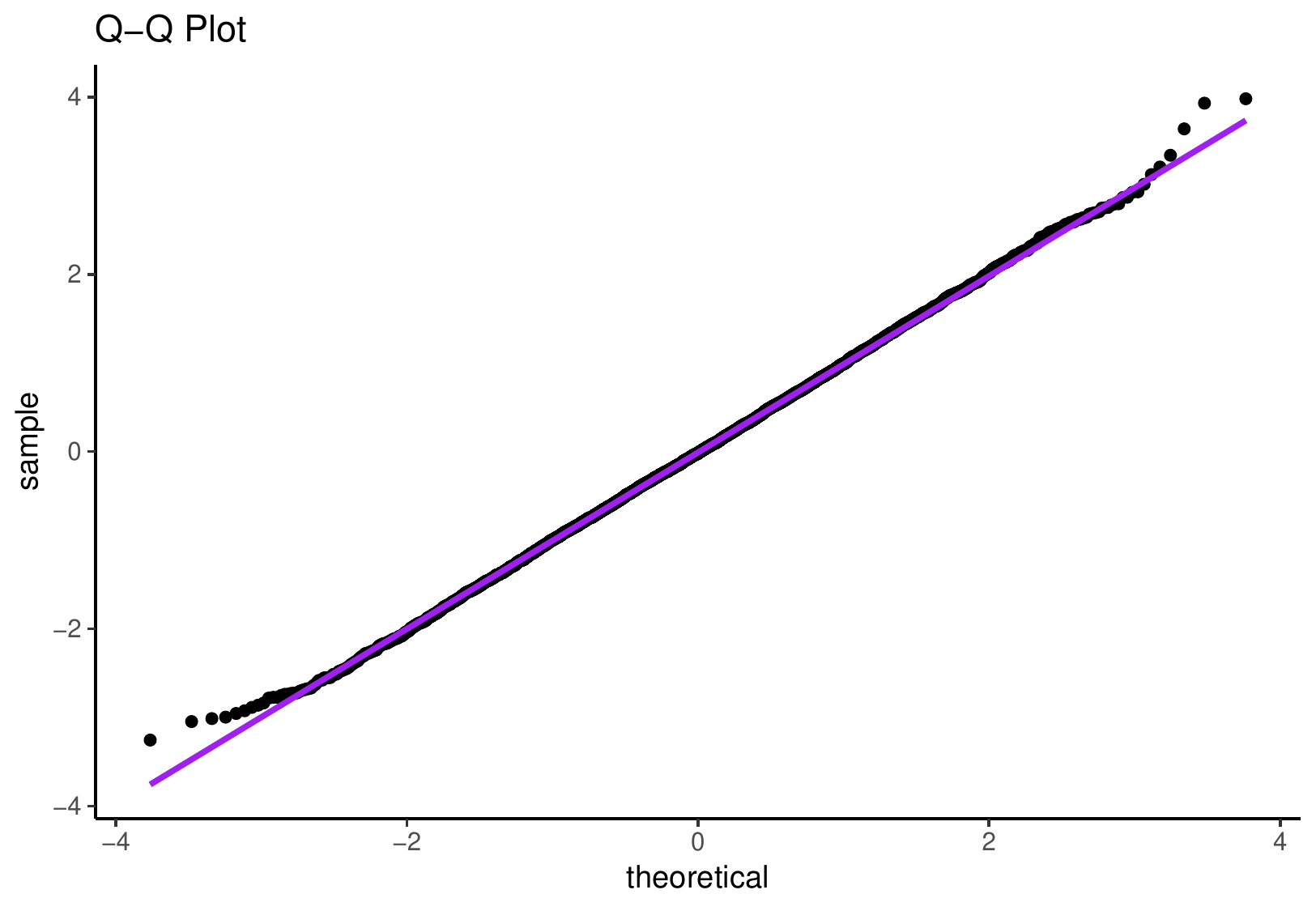} \end{center}

\subsection{Interpreting the results - Proceed with
caution}\label{interpreting-the-results---proceed-with-caution}

The application of an HMM to positional data is meant to serve as the
data generating mechanism. Ideally, the estimated states serve as
proxies for biologically meaningful behaviors, but\ldots{}

\begin{center}\rule{0.5\linewidth}{\linethickness}\end{center}

\begin{center}
\textit{Important Lesson:}\\
\vspace{3mm}
The adequacy of the HMM is in no way validated by how well the states \\ correspond to general behaviors of interest.
\end{center}

\begin{center}\rule{0.5\linewidth}{\linethickness}\end{center}

Keeping this in mind, is an HMM useful? Yes, of course! The usefulness
lies in combining biological expertise with a modeling framework that is
intuitive (from a biological standpoint) to use as the data generating
mechanism for the observed movement data. But an HMM is not magic, nor
will any other unsupervised technique magically identify general
behaviors of interest without some understanding of the biological
mechanism.

For various animals, the movement patterns that manifest themselves in
positional data do tend to follow a general pattern: directed movements
tend to correlate with large step lengths and larger turning angles are
generally associated with shorter step lengths. In terrestrial animals,
this can broadly serve as proxies for areas in which the animal will
forage or travel through. In marine animals, like sharks, we have
interepreted the states to correspond to area-restricted search and
traveling behavior. Nonetheless, the HMM is useful for clustering
movement patterns into these general behavioral states. From there, we
can incorporate covariates to understand what may drive an animal to
remain in a certain area (chum in the water, habitat quality, etc.).

En fin, an HMM can be quite a useful tool for the analysis of animal
movement data, blending important ecological knowledge with
sophisticated modeling techniques. And importantly, inferences can be
made in the Stan programming language.

\section{Acknowledgements}\label{acknowledgements}

We thank Juan M. Morales and Roland Langrock for feedback on an earlier
version.

\section{References}\label{references}

Alexandrovich, G., Holzmann, H. \& Leister, A. (2016) Nonparametric
identification and maximum likelihood estimation for hidden Markov
models. \emph{Biometrika}, \textbf{103}, 423--434.

Betancourt, M. (2017). \emph{Identifying Bayesian Mixture Models.}
Retrieved from
\url{https://betanalpha.github.io/assets/case_studies/identifying_mixture_models.html}

Betancourt, M. (2018). \emph{A Principled Bayesian Workflow}. Retrieved
from
\url{https://betanalpha.github.io/assets/case_studies/principled_bayesian_workflow.html}

Gabry, J. and Mahr, T. (2018). \emph{bayesplot: Plotting for Bayesian
Models.} R package version 1.5.0.
\url{https://CRAN.R-project.org/package=bayesplot}

Langrock, R. \& Zucchini, W. (2011) Hidden Markov models with arbitrary
state dwell-time distributions. \emph{Computational Statistics and Data
Analysis}, \textbf{55}, 715--724.

Langrock, R., Kneib, T., Sohn, A., \& DeRuiter, S. L. (2015)
Nonparametric inference in hidden Markov models using P-splines.
\emph{Biometrics}, \textbf{71}(2), 520--528.

Michelot, T., Langrock, R. \& Patterson, T.A. (2016) moveHMM: an R
package for the statistical modelling of animal movement data using
hidden Markov models. \emph{Methods in Ecology and Evolution},
\textbf{7}, 1308--1315.

Morales, J.M., Haydon, D.T., Frair, J., Holsinger, K.E., \& Fryxell,
J.M. (2004). Extracting more out of relocation data: building movement
models as mixtures of random walks. \emph{Ecology}, \textbf{85}(9),
2436--2445.

Plummer, M., Best, N., Cowles, K and Vines, K. (2006). \emph{CODA:
Convergence Diagnosis and Output Analysis for MCMC}, R News, vol 6, 7-11

Stan Development Team (2018). \emph{RStan: the R interface to Stan.} R
package version 2.17.3. \url{http://mc-stan.org/}.

Wickham, H. \emph{ggplot2: Elegant Graphics for Data Analysis.}
Springer-Verlag New York, 2016.

Zucchini, W., MacDonald, I.L. Langrock, R. (2016) \emph{Hidden Markov
Models for Time Series: An Introduction using R}, 2nd Edition, Chapman
\& Hall/CRC, FL, Boca Raton

\pagebreak

\section{Extras}\label{extras}

\subsection{State decoding}\label{state-decoding}

We can obtain inferences into the hidden state process, using either
\texttt{global\ decoding\textquotesingle{}\ (Viterbi\ algorithm)\ or}local
decoding' (forward-backward algorithm). For a set of estimated
parameters, the Viterbi algorithm computes the sequence of states most
likely to have given rise to the observed data. The forward-backward
algorithm is used to derive state probabilities, i.e.~probabilities of
being in each state at each time step.

In Stan, we can obtain the most likely state sequence (and/or state
probabilities) for each posterior draw. We include the Viterbi and
forward-backward algorithms in the `generated quantities' block, as
shown below.

\begin{verbatim}
generated quantities {
  int<lower=1,upper=N> viterbi[T];
  real stateProbs[T,N];
  vector[N] lp;
  vector[N] lp_p1;
  
  // Viterbi algorithm (most likely state sequence)
  {
    real max_logp;
    int back_ptr[T, N];
    real best_logp[T, N];
    
    for (t in 1:T) {
      if(t==1 || ID[t]!=ID[t-1]) {
        for(n in 1:N)
          best_logp[t, n] = gamma_lpdf(steps[t] | shape[n], rate[n]);
      } else {
        for (n in 1:N) {
          best_logp[t, n] = negative_infinity();
          for (j in 1:N) {
            real logp;
            logp = best_logp[t-1, j] + log_theta[t,j,n];
            if(steps[t]>0)
              logp = logp + gamma_lpdf(steps[t] | shape[n], rate[n]);
            if(angles[t]>(-pi()))
              logp = logp + von_mises_lpdf(angles[t] | loc[n], kappa[n]);
            
            if (logp > best_logp[t, n]) {
              back_ptr[t, n] = j;
              best_logp[t, n] = logp;
            }
          }
        }
      }
    }
    
    for(t0 in 1:T) {
      int t = T - t0 + 1;
      if(t==T || ID[t+1]!=ID[t]) {
        max_logp = max(best_logp[t]);
        
        for (n in 1:N)
          if (best_logp[t, n] == max_logp)
            viterbi[t] = n;
      } else {
        viterbi[t] = back_ptr[t+1, viterbi[t+1]];
      }
    }
  }
  
  // forward-backward algorithm (state probabilities)
  {
    real logalpha[T,N];
    real logbeta[T,N];
    real llk;
    
    // log alpha probabilities
    for(t in 1:T) {
      if(t==1 || ID[t]!=ID[t-1]) {
        for(n in 1:N)
          lp[n] = -log(N);
      }
      
      for (n in 1:N) {
        lp_p1[n] = log_sum_exp(to_vector(log_theta_tr[t,n]) + lp);
        if(steps[t]>=0)
          lp_p1[n] = lp_p1[n] + gamma_lpdf(steps[t] | shape[n], rate[n]);
        if(angles[t]>=(-pi())) {
          lp_p1[n] = lp_p1[n] + von_mises_lpdf(angles[t] | loc[n], kappa[n]);
        }
        logalpha[t,n] = lp_p1[n];
      }
      lp = lp_p1;
    }
    
    // log beta probabilities
    for(t0 in 1:T) {
      int t = T - t0 + 1;
      
      if(t==T || ID[t+1]!=ID[t]) {
        for(n in 1:N)
          lp_p1[n] = 0;
      } else {
        for(n in 1:N) {
          lp_p1[n] = log_sum_exp(to_vector(log_theta_tr[t+1,n]) + lp);
          if(steps[t+1]>=0)
            lp_p1[n] = lp_p1[n] + gamma_lpdf(steps[t+1] | shape[n], rate[n]);
          if(angles[t+1]>=(-pi()))
            lp_p1[n] = lp_p1[n] + von_mises_lpdf(angles[t+1] | loc[n], kappa[n]);
        }
      }
      lp = lp_p1;
      for(n in 1:N)
        logbeta[t,n] = lp[n];
    }
    
    // state probabilities
    for(t0 in 1:T) {
      int t = T - t0 + 1;
      if(t==T || ID[t+1]!=ID[t])
        llk = log_sum_exp(logalpha[t]);
      for(n in 1:N)
        stateProbs[t,n] = exp(logalpha[t,n] + logbeta[t,n] - llk);
    }
  }
}
\end{verbatim}

\end{document}